\documentclass[abstract]{aa}
\usepackage[T1]{fontenc}
\usepackage{ae,aecompl}
\usepackage{graphicx}
\usepackage{txfonts}
\usepackage{amsmath}
\usepackage{amssymb}
\usepackage{sidecap}
\usepackage{xcolor}
\usepackage{subfig}
\usepackage{appendix}
\usepackage{rotating}
\usepackage{tablefootnote}
\usepackage{comment}
\usepackage{booktabs,caption}
\usepackage[flushleft]{threeparttable}
\usepackage{multirow}
\usepackage{hyperref}
\usepackage{upgreek}
\usepackage[switch]{lineno}

\bibpunct{(}{)}{;}{a}{}{,} 

\def \msun{\rm \, M_\odot}

\hypersetup{
  colorlinks   = true, 
  urlcolor     = blue, 
  linkcolor    = red, 
  citecolor    = blue 
}

\titlerunning{The radio galaxy NGC 6086} 
\authorrunning{S. Candini, M. Brienza, A. Bonafede et al.}

\begin{document} 

\title{New filamentary remnant radio emission and duty\\ cycle constraints in the radio galaxy NGC 6086} 

\author{S. Candini\inst{1},
    M. Brienza\inst{2,1}, 
    A. Bonafede\inst{1,3},
    K. Rajpurohit\inst{6,1,3},
    N. Biava\inst{1,3},
    M. Murgia\inst{4},\\
    F. Loi\inst{4},
    R. J. van Weeren\inst{5},
    F. Vazza\inst{1,3,7}
    }

\institute{Dipartimento di Fisica e Astronomia, Università di Bologna, via P. Gobetti 93/2, I-40129, Bologna, Italy
    \and 
    INAF - Osservatorio di Astrofisica e Scienza dello Spazio di Bologna, via Gobetti 93/3, I-40129 Bologna, Italy
    \and
    INAF - Istituto di Radioastronomia, Bologna Via Gobetti 101, I-40129 Bologna, Italy
    \and
    INAF - Osservatorio Astronomico di Cagliari, Via della Scienza 5, I-09047 Selargius (CA), Italy
    \and
    Leiden Observatory, Leiden University, PO Box 9513, 2300 RA Leiden, The Netherlands
    \and
    Harvard-Smithsonian Center for Astrophysics, 60 Garden Street, Cambridge, MA 02138, USA
    \and 
    Universit\"at Hamburg, Hamburger Sternwarte, Gojenbergsweg 112, 21029, Hamburg, Germany
    }
\date{Accepted 21 June 2023; received 30 May 2023; in original form \today}

\abstract{Radio galaxies are a subclass of active galactic nuclei in which accretion onto the supermassive black hole releases energy into the environment via relativistic jets. The jets are not constantly active throughout the life of the host galaxy and alternate between active and quiescent phases. Remnant radio galaxies are detected during a quiescent phase and define a class of unique sources to constrain the AGN duty cycle.

We present, for the first time, a spatially resolved radio analysis of the radio galaxy associated with the galaxy NGC 6086 down to 144 MHz and constraints on the spectral age of the diffuse emission to investigate the duty cycle and evolution of the source.

We use three new low-frequency, high-sensitivity observations, the first one is performed with the Low Frequency Array at 144 MHz and the other two are performed with the upgraded Giant Metrewave Radio Telescope at 400 MHz and 675 MHz, respectively. To these, we add two Very Large Array archival observations at higher frequencies (1400 and 4700 MHz).

In the new observations in the frequency range 144-675 MHz we detect a second pair of larger lobes and three regions within the remnant emission with a filamentary morphology. We analyse the spectral index trend in the inner remnant lobes and see systematic steeper values ($\rm{\alpha_{low}\sim}$1.1-1.3) at the lower frequencies compared to the GHz frequencies ($\rm{\alpha_{high}\sim}$0.8-0.9). Steeper spectral indices are found in the newly detected outer lobes (up to $\rm{\alpha_{outer}\sim}$2.1), as expected if they trace a previous phase of activity of the AGN. However, the differences between the spectra of the two outer lobes suggest different dynamical evolution within the intragroup medium during their expansion and/or different magnetic field values.

Using a single-injection radiative model and assuming equipartition conditions, we place constraints on the age of the inner and outer lobes and derive the duty cycle of the source. We estimate that the duration of the two active phases was 45 Myr and 18 Myr and the duration of the two inactive phases was 66 Myr and 33 Myr. This results in a total active time of $\rm{t_{on}\sim39 \%}$. 

The filamentary structures have a steep spectral index ($\sim1$) without any spectral index trend and only one of them shows a steepening in the spectrum. Their origin is not yet clear, but they may have formed due to the compression of the plasma or due to magnetic field substructures.
}

\keywords{radiation mechanisms: non-thermal - radio continuum: galaxies - galaxies: active - galaxies: groups: general - galaxies: individual: NGC 6086 }

\maketitle
\section{Introduction}
\label{sec:intro}
Radio Galaxies are a peculiar class of galaxies with luminosities in the radio band up to $10^{46}\ \rm{erg/s}$ between $10\ \rm{MHz}$ and $100\ \rm{GHz}$. These galaxies are a subclass of active galactic nuclei (AGN), which are the elliptical galaxies that host an active supermassive black hole (SMBH) in the nuclear region. The central SMBH can give rise to a couple of relativistic jets which drag off relativistic plasma and magnetic field.

The relativistic particles lose energy through synchrotron radiation and Inverse Compton (IC) scattering. The former process is best observed at radio wavelengths, while the latter has the peak of emission in the X-ray band. The AGN-driven jets can have a range of sizes, from less than a kpc \citep{odea2021} up to hundreds of kpc, or even a few Mpc in the case of giant radio galaxies \citep[e.g.][]{willis1974, saripalli1986, dabhade2020, gurkan2022, oei2022}.

The importance of studying radio galaxies is not limited to understanding the physics of these sources. Indeed, the energy injected by jets in the environment has a crucial impact on the evolution of the host galaxies and the external intra-group medium (IGrM) or intra-cluster medium (ICM) \citep{hardcastle2020}. 
One essential ingredient to quantify the impact of jets in their surroundings is their duty cycle, that is the fraction of time during which the sources are active \citep{romano2014}. 

Statistical studies on radio source populations suggest that this is strongly dependent on the mass of the host galaxy. \cite{sabater2019}, for example, found that in the massive galaxies ($\msun>10^{11}$) the radio AGN activity is always switched on at some level.

The life cycle of the AGN is driven by the accretion of matter into the central SMBH. Indeed, the cycle is characterised by the succession of active phases and quiescent ones with variable timescales. To characterise and understand the central engines of the AGN, it is crucial to study the duty cycle of the sources. 


The best sources where the jet duty cycle can be directly probed are restarted radio galaxies, which are galaxies showing recurrent jet activity and multiple generations of radio emission \citep[e.g.][]{venturi2004, marecki2006, jamrozy2007, parma2007, mahatma2018, biava2021, schellenberger2021, Ubertosi2021}. The clearest examples of this recurrent activity are the double-double radio galaxies \citep[e.g.][]{schoenmakers2000, konar2006, orru2010}, this class of galaxies clearly shows two generations of lobes inflated at different times. 
However, there are also cases where more than two phases of activity have been reported, For example, in the case of Fornax A, a recent spatially resolved spectral study has suggested that it is possible to distinguish at least three distinct phases of activity in the past history of the radio source \citep{maccagni2020}. A few other sources with three detected active phases have been found in the last few years with radio observation \citep{brocksopp2007, singh2016} or with X-ray observations which highlight three cavities in the ICM \citep{randall2015}.

Particularly interesting is to study the jet duty cycle in small systems such as galaxy groups. Here, the energy released by the jets is comparable to, or higher than, the binding energy of the group itself, forcing baryon depletion \citep{lagana2013, kolokythas2019}. For this reason, the knowledge of the duty cycle and the energy released by AGN are also essential to the galaxy evolutionary models, large-scale system simulations and for constraining cosmological models \citep[see][for a full review of AGN feedback in galaxy groups]{eckert2021}.

In this paper, we present deep, low-frequency radio observations performed with the LOw-Frequency ARray (LOFAR, \citealp{vanhaarlem2013}) and the upgraded Giant Metrewave Radio Telescope (uGMRT, \citealp{gupta2017}) of the radio galaxy B2 1610+29, which is associated with the galaxy NGC 6086, located at the centre of the galaxy group Abell 2162. Alongside the new observations, we use Very Large Array (VLA) archival radio observations of the source at higher frequencies to perform spectral analysis and give constraints on the duty cycle of the jet activity.

\begin{figure}[!htp]
\includegraphics[width=0.48\textwidth]{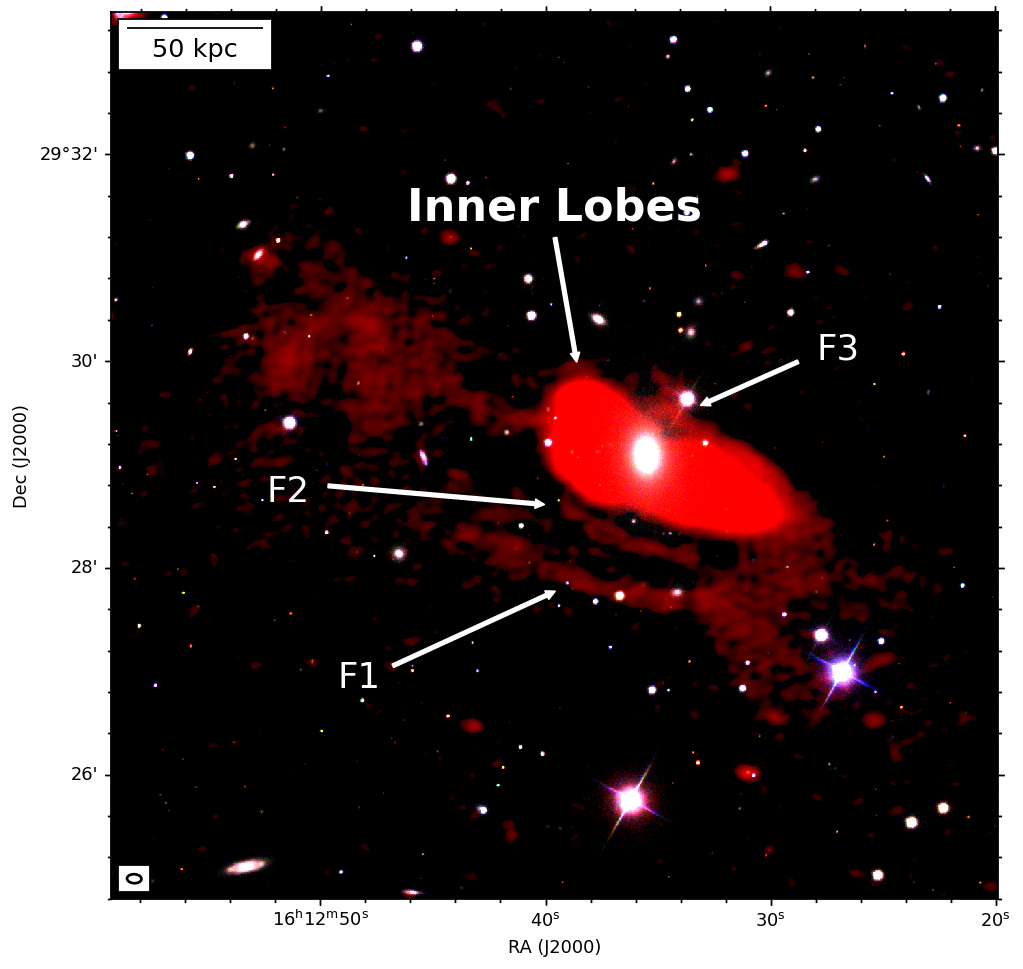}
\caption{Composite image of the radio galaxy NGC 6086, obtained using optical (SDSS r-band, g-band and i-band) and radio (LOFAR at 144 MHz with 5.22 arcsec $\times$ 8.37 arcsec resolution) images. The most significant features of the system are labelled, the beam is shown in the bottom-left corner and a reference physical scale is shown in the top-left one.}
\label{fig:rgb}
\end{figure}

The outline of the paper is as follows. In Sect. \ref{sec:overview}, we provide an overview of NGC 6086. Radio observations and data reduction procedures are described in Sect. \ref{sec:data}. In Sect. \ref{sec:analysis}, we present the results of our analysis of the radio emission over different spatial scales. In Sect. \ref{sec:age} and \ref{sec:duty_cycle}, we present the radiative age and the duty cycle estimates of the source, respectively. The discussion on the results of NGC 6086 is reported in Sect. \ref{sec:discussion} and a summary of our main findings is reported in Sect. \ref{sec:concl}. The cosmology adopted throughout the paper assumes a flat universe with the following parameters: $\rm H_{0} = 70$ $\rm km$ $\rm s^{-1}$ $\rm Mpc^{-1}$, $\rm \Omega_{\Lambda} =0.7$, $\rm \Omega_{M} =0.3$ \citep{giacintucci2007}; at the redshift of NGC 6086, z=0.0318 \citep{murgia2011}, 1 arcsec corresponds to 0.635 kpc; the synchrotron power-law distribution is defined as $\rm{S_{\nu}\propto\nu^{-\alpha}}$. 

\section{Overview of NGC 6086}
\label{sec:overview}

\begin{figure}[!htp]
\includegraphics[width=0.48\textwidth]{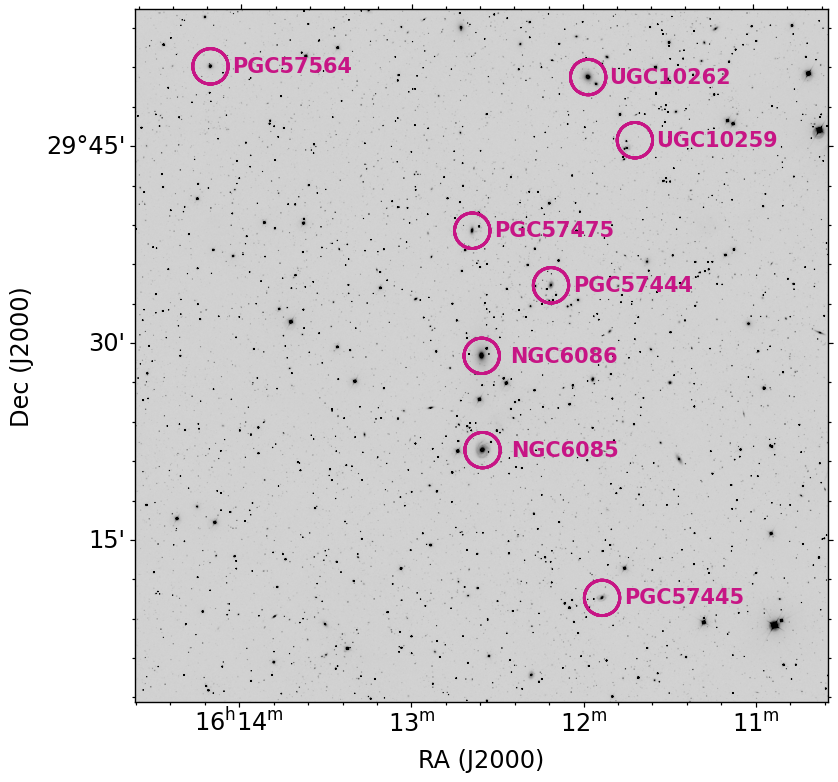}
\caption{SDSS r-band image of the galaxy group Abell 2162. Bordeaux circles mark the main galaxies of the group.}
\label{fig:6086sdss}
\end{figure}

NGC 6086 is the central galaxy of the galaxy group Abell 2162, which is one of the 37 members \citep{abell1989} of the Hercules supercluster \citep{einasto2001} (see Fig. \ref{fig:6086sdss}). The X-ray luminosity of the system in the 0.3-4.5 kev band is 3.1$\rm \times 10^{42}$ erg/s based on Einstein observations \citep{burns1994}, while in the 0.5–2.0 keV band is 1.9$\rm \times 10^{42}$ erg/s 
based on ROSAT observations \citep[][rescaled for the cosmological model adopted in this work]{mahdavi1997}. 

NGC 6086 is located close to the peak of the X-ray emission, however, no morphological analysis of X-ray emission of the surrounding IGrM is available in the literature. The galaxy is an elliptical and it hosts a low-power ($\rm P_{151MHz}=1.4\times 10^{24}$ W/Hz) radio galaxy (B2 1610+29) with a total extension of $\sim$100 kpc $\times$ 30 kpc \citep{giacintucci2007, murgia2011}. In Fig. \ref{fig:rgb} we show the new LOFAR radio image of the source presented in this work overlaid on the optical image. The morphology of the radio source has been analysed in the literature (with GMRT and VLA observations) and consists of two low surface brightness and relaxed lobes sitting on the opposite side of the host galaxy, with no indications of compact components such as jets, hotspots or active core \citep{parma1986, owen1997}. The lack of nuclear radio emission was further confirmed by Very Long Baseline Array data at 5 GHz presented by \cite{liuzzo2010}. These morphological properties, combined with a very curved radio spectrum in the range 74-8350 MHz \citep{murgia2011}, suggest that the nuclear activity is currently switched off and that the lobes represent the remnants of a past phase of jet activity. Based on spectral ageing models \citep{komissarov1994}, \cite{murgia2011} estimated that the total age of the plasma is in the range of 55-60 Myr and the jets switched off 25-45 Myr ago. In this work, we present new observations with the LOFAR and uGMRT interferometers to further investigate the nature of the radio galaxy and give constraints to its duty cycle.

\section{Observations and data processing}
\label{sec:data}
\subsection{LOFAR 144 MHz}
\label{lofar}

The target has been observed by LOFAR with the High Frequency Antennas at a central frequency of 144 MHz as part of LoTSS (LOFAR Two-metre Sky Survey, \citep{shimwell2017}). We used the two 8-hour datasets related to the LoTSS pointings P243+30 and P242+28, whose centres lie at 0.54 and 2.06 degrees away from the target position, respectively. Both datasets have been recorded following the observational setup of LoTSS. The total bandwidth of the observations is 48 MHz, covering frequencies in the range $120-168\ $MHz and the frequency resolution is 12.2 kHz. All four polarisations were recorded (XX, XY, YX, YY) with a sampling time equal to 1s. The primary calibrator was observed at the end of the observing run for 10 minutes. The full LOFAR array has been used for the observations but for this work, we only analyse the data collected by the Dutch stations (baselines $\leq$100 km).

Before being stored in the LOFAR long-term archive, the observatory flagged the data for radio interference and then averaged them by a factor of 4 in frequency. The data were then processed using the standard LoTSS procedures. An outline of the data processing is presented in this section, and we refer the reader to \cite{tasse2021} and \cite{shimwell2022} for a detailed description of the procedure.

The PreFactor pipeline\footnote{\url{https://github.com/lofar-astron/prefactor}} \citep{vanweeren2016, williams2016} was used to correct the data for direction independent effects such as offsets between XX and YY phases, ionospheric Faraday rotation and clock offsets (see \citealt{degasperin2019}). 

The DDF-pipeline\footnote{\url{https://github.com/mhardcastle/ddf-pipeline}} v2.2 was used to perform a direction-dependent self-calibration, to correct for distortions of the ionosphere and errors in the beam model.

This pipeline is described in \cite{shimwell2019} and \cite{tasse2021}, it uses kMS (\citealt{tasse2014} and \citealt{smirnov2015}) to derive direction-dependent calibration solutions and DDFacet for imaging (\citealt{tasse2018}). To further improve the quality of the calibration and facilitate data imaging, all sources outside a square region of $\sim$30-arcmin side and centred on the target were subtracted from the UV-data, and final loops of self-calibration were performed (see \citealp{vanweeren2021} for a detailed description of the procedure). Baselines below 80$\rm \lambda$ were not considered.

The final images were produced using multiscale cleaning in WSClean (version 2.8, \citealp{offringa2014}) and are presented in Fig. \ref{fig:6086}. The different sets of imaging parameters were chosen to both recover the small morphological features and enhance the large-scale diffuse emission. The parameters used together with the final image properties are listed in Table \ref{tab:imageparam}.

\subsection{uGMRT 400 MHz and 675 MHz}

We observed the target with the uGMRT in September 2020, in both band-3 (300-500 MHz) and band-4 (550-950 MHz) (see Table \ref{tab:data}). The total on-source observing time was $\sim$6.3 hours for band-3 and $\sim$7 hours for band-4 observation. The flux density calibrator used was 3C48 and it was observed for 8 minutes at the beginning or the end of each observing run. In both frequency ranges, the total bandwidth was divided into 4096 channels and the integration time-step was set to 5.3 seconds.

We calibrated the data using the SPAM pipeline \citep{intema2014, intema2017} upgraded for handling new wideband uGMRT data following the standard procedure\footnote{\url{http://www.intema.nl/doku.php?id=huibintemaspampipeline}} and we set the absolute flux density scale according to \cite{scaife2012}. Due to severe radio frequency interference, data above 850 MHz in band-4 were removed. Using the output calibrated data, we created the final images using multiscale cleaning with WSClean. 
As for the LOFAR data, we imaged both datasets with two sets of imaging parameters reported in Table \ref{tab:imageparam} together with the final image properties. 

\subsection{VLA 1.4 GHz and 4700 MHz}
We selected archival VLA observations to expand the frequency coverage of the source above 675 MHz, as needed for a more detailed spectral analysis. We selected two datasets where NGC 6086 was observed at 1.4 GHz with B and C array configurations and a second dataset at 4.7 GHz with the D array configuration. The details of the observations used are reported in Table \ref{tab:data}.

All the datasets were reduced with Common Astronomy Software Applications (\texttt{CASA} version 5.5.0-149; \citealp{mcmullin2007}). After manual flagging, we performed the standard calibration, with the flux density scale set according to \cite{perley2013} for all datasets. We calibrated the two datasets at 1.4 GHz singularly, using as primary calibrator 3C286 (or 1328+307) for both. We used the source 1219+285 as the phase calibrator which is the closest in time and position for the C array configuration dataset, while we chose the source 1607+268 for the B array configuration one. We performed three loops of phase self-calibration on the C configuration data to refine the gain solutions. The final image has a resolution of 17.4 arcsec $\times$ 13.8 arcsec and a noise equal to $\rm{350\ \mu Jy/beam}$.
On the B configuration data, we performed two rounds of phase self-calibration and we obtained a final image with 6.2 arcsec $\times$ 4.6 arcsec and a noise equal to $\rm{200\ \mu Jy/beam}$.
We then combined the two calibrated datasets at 1.4 GHz to improve the UV-coverage. We performed a few loops of phase self-calibration and we made a final imaging imposing a restoring beam of 14 arcsec to match the highest resolution available for the 4.7 GHz image. The rms noise level of the combined image at 1.4 GHz is $\rm{120 \mu Jy/beam}$. 

For the dataset at 4.7 GHz in D configuration, we followed the same calibration and imaging procedure described above. The resolution of the final image is 14.5 arcsec $\times$ 12.9 arcsec and the rms noise level is $\rm{100\ \mu Jy/b}$.
A summary of the radio image properties is reported in Table \ref{tab:imageparam} and the two final VLA images at 14 arcsec of resolution are shown in Appendix \ref{Fig:ngc6086_vla}.

\begin{table}[!htp]
\begin{threeparttable}
    \caption{Summary of the observations of NGC 6086 used in this work.}
    \begin{tabular}{c c c c}
        \hline\hline
        \multirow{2}{*}{Telescope} & Frequency & TOS$^1$ & Date \\
        & [MHz] & [hh:mm] & [dd/mm/yy]\\
        \hline
        \multirow{2}{*}{LOFAR HBA$^2$} & \multirow{2}{*}{120-168} & \multirow{2}{*}{16:00} & 14/05/18\\
        &&& 08/11/19\\
        uGMRT & 300-500 & 06:20 & 06/09/20\\
        uGMRT & 550-950 & 07:00 & 22/09/20\\
        VLA (B array)$^3$ & 1400 & 00:12 & 21/02/93\\
        VLA (C array)$^3$ & 1400 & 00:07 & 29/01/91\\
        VLA (D array)$^3$ & 4700 & 00:10 & 01/10/00\\
        \hline\hline  
    \end{tabular}              
\begin{tablenotes}
    \item $^1$Time On Source.
    \item $^2$Observations not centered on the target (see Sect. \ref{lofar} for details).
    \item $^3$Archival observation.
\end{tablenotes}
     \label{tab:data}
\end{threeparttable}
\end{table}

\begin{table*}[!htp]
    \caption{Summary of the radio images of NGC 6086.}
    \centering
        \begin{tabular}{c c c c c }
        \hline\hline
        Central frequency & Beam & \multirow{2}{*}{Weighting} & UV-taper & RMS noise\\
        $[\rm{MHz}]$ & [arcsec$\times$arcsec]& & [arcsec] & [mJy/beam]\\
        \hline
        144 & 8.4$\times$5.2 & Briggs -0.5 & - & 0.11\\
        144 & 28.3$\times$24.6 & Briggs -0.5 & 20 & 0.2 \\
        400 & 8$\times$5.5 & Briggs 0 & - & 0.035 \\
        400 & 31$\times$27 & Briggs 0.5 & 25 & 0.14 \\
        675 & 4.8$\times$3.7 & Briggs 0 & - & 0.014 \\
        675 & 26$\times$17 & Briggs 0.5 & 25 & 0.08 \\
        1400 (B) & 6.2$\times$4.6 & Natural & - & 0.2 \\
        1400 (C) & 17.4$\times$13.8 & Briggs 0 & - & 0.35 \\
        4700 & 14.5$\times$12.9 & Uniform & - & 0.1 \\
        \hline\hline  
    \end{tabular}
    \label{tab:imageparam}
\end{table*}

\section{Results}
\label{sec:analysis}

\subsection{Radio morphology}
\label{sec:radio}
\begin{figure*}[!htp]
    \centering
    \vspace{.5cm}
    \minipage{0.48\textwidth}
        \includegraphics[width=1\textwidth]{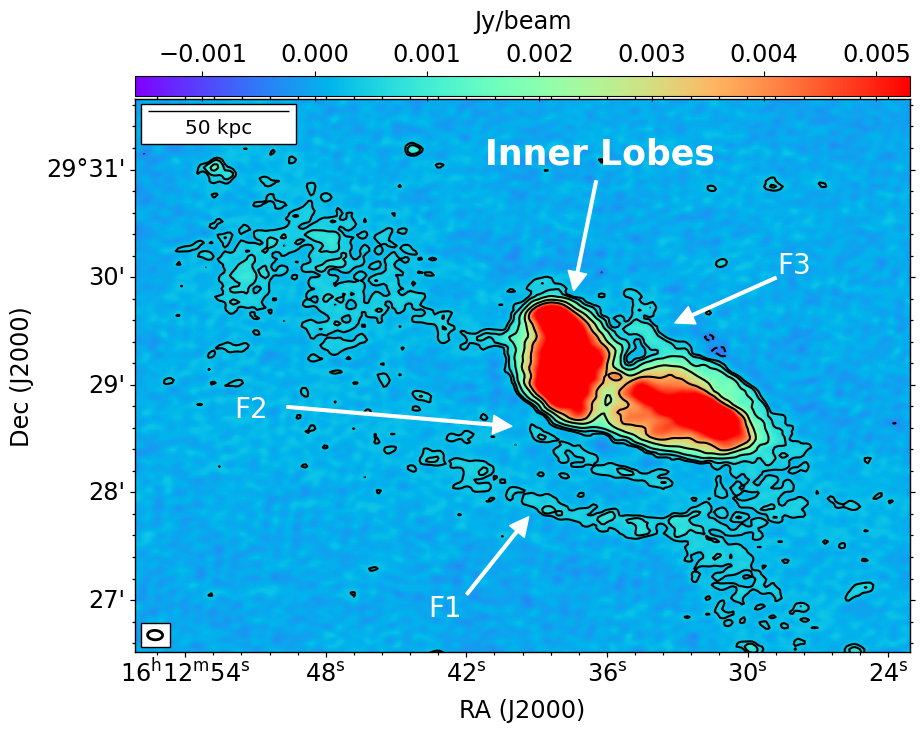}
        \endminipage\hfill
    \centering
    \minipage{0.48\textwidth}
        \includegraphics[width=1\textwidth]{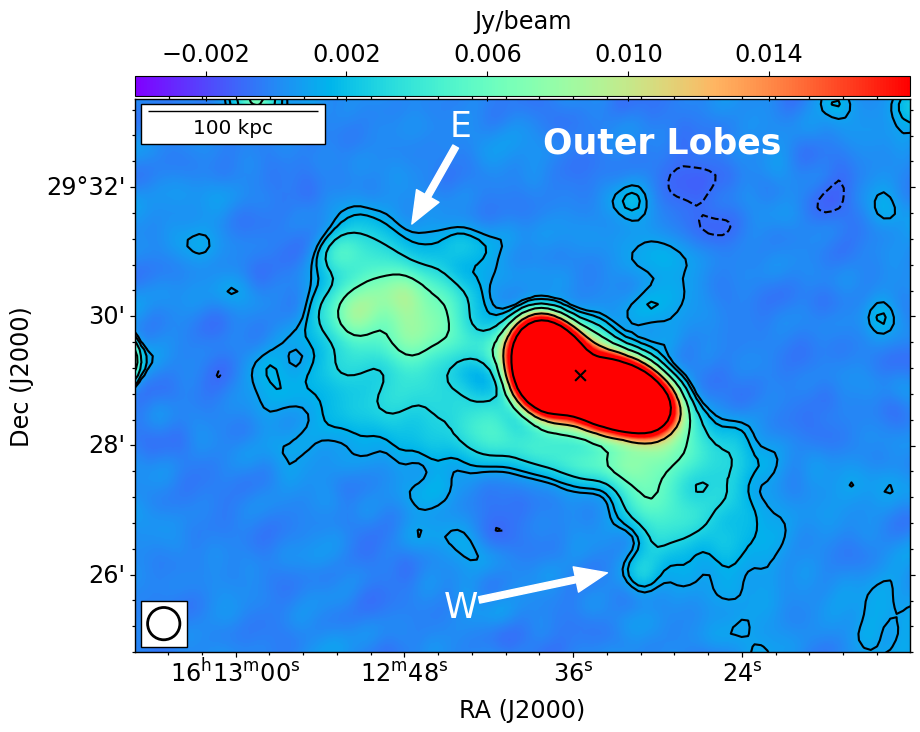}
    \endminipage\hfill
    \centering
    \minipage{0.48\textwidth}
        \includegraphics[width=1\textwidth]{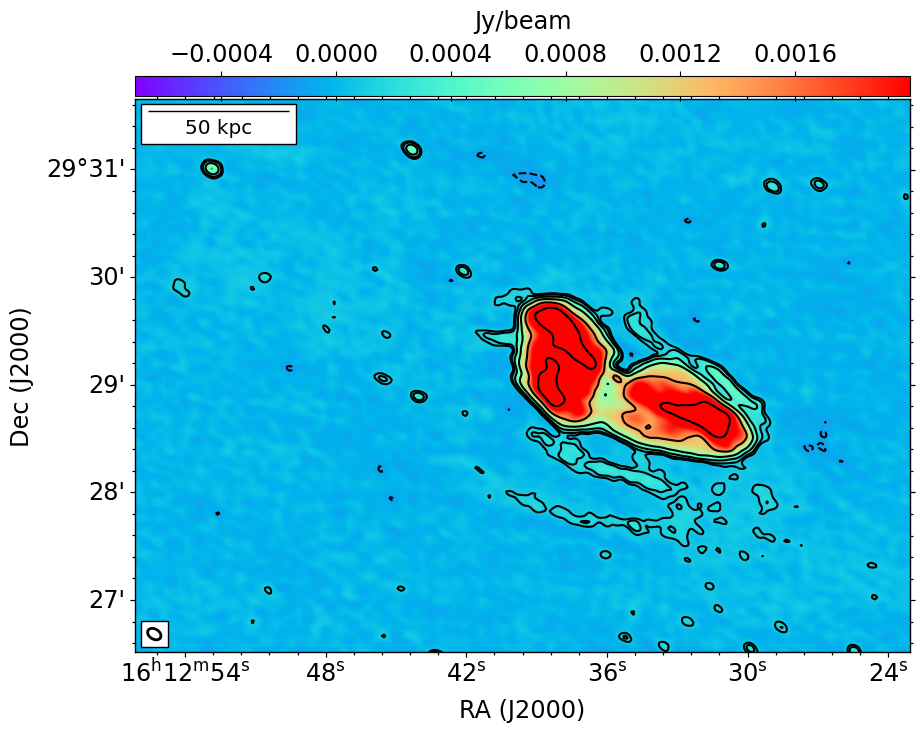}
    \endminipage\hfill
    \centering
    \minipage{0.48\textwidth}
        \includegraphics[width=1\textwidth]{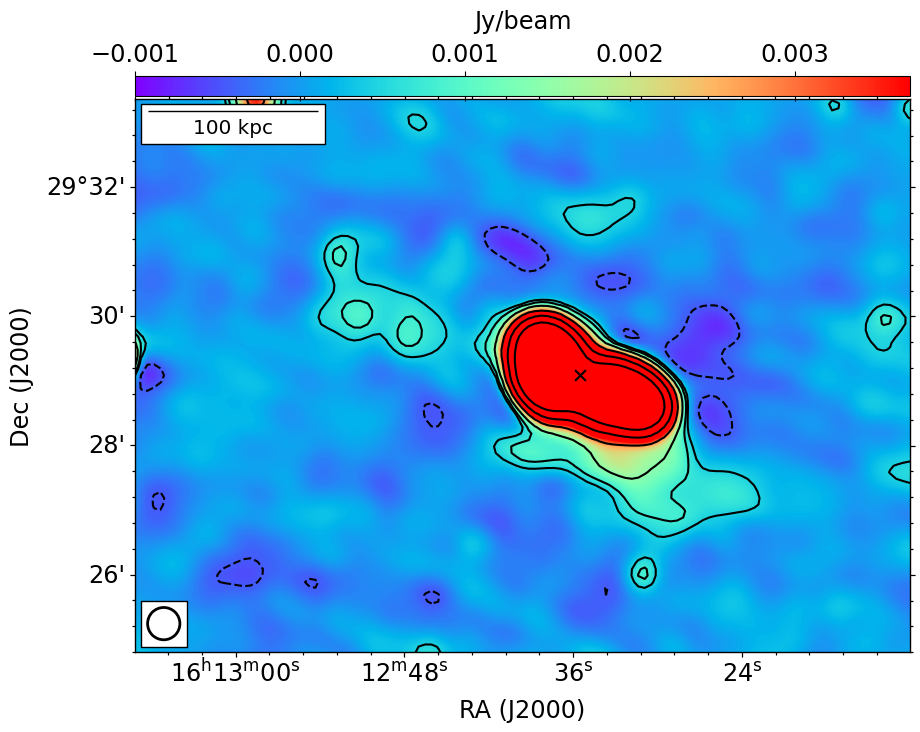}
    \endminipage\hfill
    \centering
    \minipage{0.48\textwidth}                          \includegraphics[width=1\textwidth]{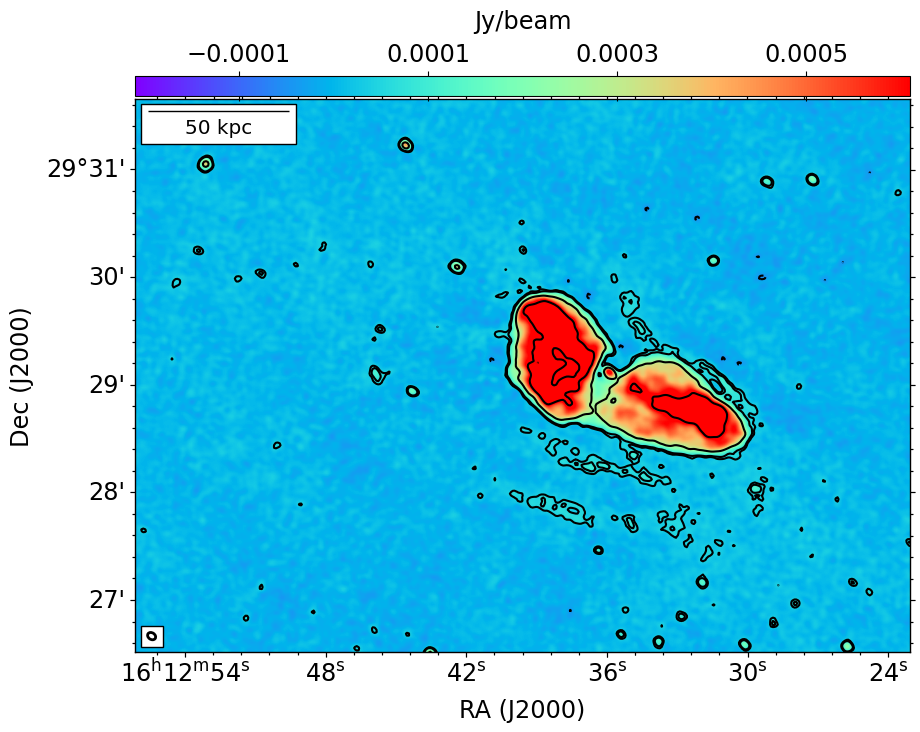}
    \endminipage\hfill
    \centering
    \minipage{0.48\textwidth}
        \includegraphics[width=1\textwidth]{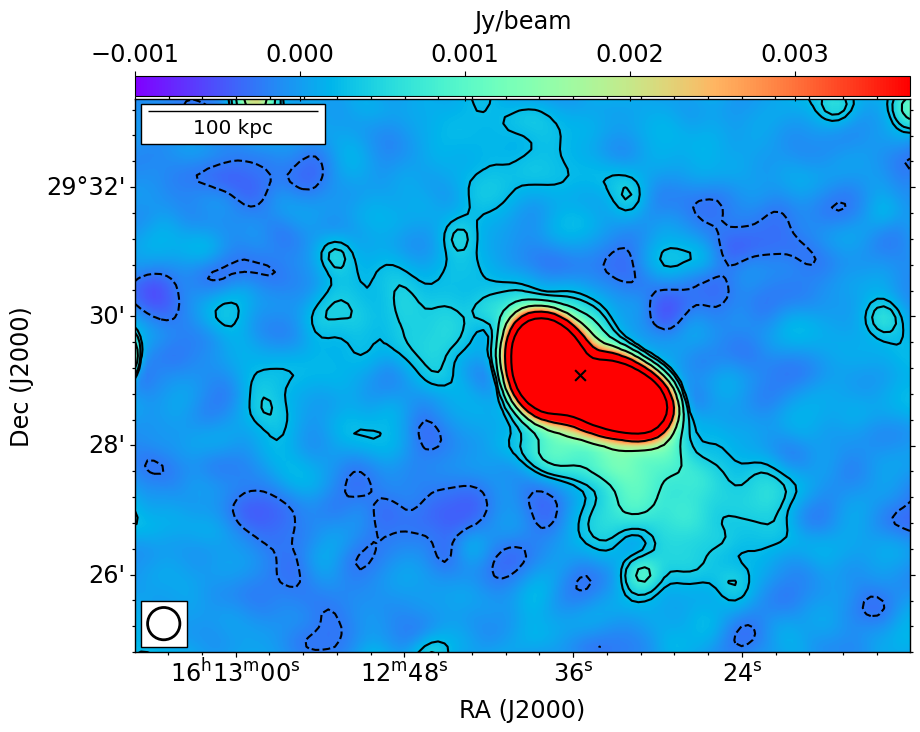}
    \endminipage\hfill
    \caption{Radio images of the source NGC 6086 at 144 MHz (top row), 400 MHz (central row) and 675 MHz (bottom row). The left column shows images at the highest resolution (8.37 $\times$ 5.22, 8 $\times $5.5, 4.8 $\times $3.7 arcsec, from the top to the bottom panel respectively)  and the right column at low resolution (30 arcsec). Contours are drawn at (-3, 3, 5, 10, 20, 30) $\times\ \sigma$ in the first panel; (-3, 3, 5, 20, 50,) $\times\ \sigma$ in the fifth panel; (-3, 3, 5, 10, 20, 40, 80)$\times\ \sigma$ in the other panels. The `$\times$' markers in the panels in the right column indicate the position of the host galaxy. The beam is shown in the bottom-left corner and a reference physical scale is shown in the top-left one.}
\label{fig:6086}
\end{figure*}

Thanks to our new low-frequency, high-sensitivity and broad-band observations in the frequency range 144-675 MHz, we are able to reveal much more extended emission than what was previously detected in NGC 6086. In Fig. \ref{fig:6086}, we show two maps for each of the three frequencies available below 1 GHz, one at high resolution (left column) and one at low resolution (right column). We note that our new 675 MHz (uGMRT, band-4) image is a factor ten deeper than the previous observation presented by \cite{giacintucci2007} at 610 MHz with the GMRT and comparable beam. Our LOFAR image is the first one below 600 MHz where the combination of resolution and sensitivity allowed us to see the radio galaxy resolved.

In the new LOFAR 144 MHz high-sensitivity images (see Fig. \ref{fig:6086}, first row), we clearly detect the two central lobes already observed in the other works that have analysed NGC 6086. Around the lobes, which are distinguishable in both the high-resolution and low-resolution images, we find new, previously undetected, diffuse emission. 
In the high-resolution image (Fig. \ref{fig:6086}, top row left column), we can appreciate the tiniest substructures of the radio plasma. In particular, the newly detected emission appears very patchy and three filaments are clearly revealed, which we label as F1, F2, and F3. In addition, a small filament seems to emerge from the eastern lobe in the direction of the diffuse, extended emission on the East. The most prominent filament, labelled as F1, is connected to the lobe in an arc-shaped structure. F1 and F2 are almost parallel and located below the inner lobes, while F3 is located above, at least in projection. The latter is partially superimposed to the western inner lobe and this makes its analysis very limited. We measure that the extension of the filaments in the LOFAR 144 MHz image at the highest resolution available is 58 kpc $\times$ 6 kpc and 54 kpc $\times$ 6 kpc, for F1 and F2 respectively. For F3, which is superimposed to the western lobe, we get a lower limit on dimensions equal to 22 kpc $\times$ 10 kpc. 

The low-resolution image (Fig. \ref{fig:6086}, top row, right column) allows us to appreciate the full extension of the newly-discovered emission in the north-east and south-west directions.
We use the 3$\sigma$ contours as a reference to measure the full extension of NGC 6086 which is equal to $\sim$7.3 arcmin and corresponds to $\sim$280 kpc. In particular, the eastern region reaches a distance from the host galaxy of $\sim$165 kpc and it is more extensive than the western one, which is only detected up to $\sim$111 kpc from it. 

Because of the observed morphology, we suggest that the newly detected, large-scale emission might represent old AGN remnant plasma produced during a past phase of jet activity of NGC 6086. In the rest of the paper, we will refer to the more extended, newly-discovered structures as `outer lobes', while we refer to the brighter, central structures as `inner lobes'.
Both inner and outer lobes are detected in the uGMRT images (see Fig. \ref{fig:6086} middle and bottom panels), even though not to the full extent of the 144 MHz images. The filaments are clearly distinguishable in the high-resolution images at both 400 MHz and 675 MHz, while the large-scale emission is almost completely undetected. Furthermore, thanks to the higher resolution of the uGMRT images, we can undoubtedly characterise F3 as a separate structure from the inner western lobe. 
The large-scale emission is only marginally detected likely because of their very steep spectrum and low-surface brightness. 
As already found by \cite{murgia2011} in the VLA images at 1.4 GHz and 4.7 GHz only the central lobes are detected (see Appendix \ref{Fig:ngc6086_vla}). No other diffuse structures are recovered in these observations due to the sensitivity to steep spectrum emission.

Finally, we note that hints of a radio core are detected in all the low-frequency images of NGC 6086 for the first time. We measure the flux density of the core inside a beam-size region in the three images below 1 GHz. The spectrum is consistent with a power-law with $\rm{\alpha_{144MHZ}^{675MHz}}\sim1.27$. Likely for this reason, even without considering the synchrotron steepening, no sign of the radio core is found in both the VLA images because of the sensitivity of the observations. The beam-size region is $\sim3$kpc wide and the core emission is contaminated by the emission on larger scales. 
New observations at arcsec or subarcsec in the GHz regime are needed to prove whether the emission is still visible up to these frequencies or not \citep{jurlin2021} and new images at subarcesec resolution at low frequency are needed to confirm the measured flux density and thus the spectral index.


\subsection{Images sets}
\label{sec:images}
To investigate the source, we have five frequencies available in the range between 144-4700 MHz with different UV-ranges.
The UV-range is 0.1-40 $\rm{k\lambda}$ for the LOFAR dataset, 0.2-33 $\rm{k\lambda}$ for the 400 MHz and 0.3-58 $\rm{k\lambda}$ for the 675 MHz datasets. The VLA 1400 MHz (B+C configuration) has a 0.3-50.4 $\rm{k\lambda}$ range and the VLA 4700 MHz dataset has a 0.6-17 $\rm{k\lambda}$ range. To make sure we recover the flux density on the same maximum spatial scales at all observed frequencies, we exclude baselines below  300 $\rm{\lambda}$, which corresponds to $\sim$14 arcmin, much larger anyway than the target extension. 
We create the three following sets of images with different angular resolutions in order to perform the best analysis of the structures on different physical scales:
\begin{itemize}
    \item high resolution: the first set of images has an angular resolution of 7 arcsec, corresponding to 4.4 kpc. They are sensitive to the smallest-scale features of the source and the tiniest details. This set is used to investigate the properties of the filaments and to characterise their spectral shape. 
    \item mid resolution: the second set has an angular resolution of 14 arcsec, corresponding to 8.9 kpc. This is the highest resolution achieved with the 4.7 GHz (D array) and is used to perform a resolved and integrated spectral analysis of the inner lobes over the full frequency range. 
    \item low resolution: the last set of images has an angular resolution of 30 arcsec, corresponding to 19 kpc. They are sensitive to the emission on the largest scale and so we use them to investigate the emission of the outer lobes by performing a resolved and integrated analysis.
\end{itemize}

Before the spectral analysis can be performed, it is important to spatially align the radio images. Indeed, imaging and self-calibration can introduce small phase shifts that can compromise the quality of the spectral analysis (i.e. leading to unreliable spectral index values) and should be corrected. This method consists of selecting a bright point source located close to the target (in all the images to be matched) and fitting it with a 2D-Gaussian function. We select a reference image and chose the point-like source, then we shift the other images to have a matched position for the selected source.

\subsection{Spectral analysis}
\label{sec:spec}
At their injection, the relativistic particles emit energy through synchrotron radiation and have an energy distribution that follows $\rm{N(E,t)=N_0E^{-\delta}}$ (where $\delta$ is the particle energy power index). The power-law distribution lies on the assumption that particles are accelerated by Fermi mechanisms and this is consistent with the observed spectra. The energy distribution translates in the observed synchrotron power-law distribution that follows $S \propto \nu ^{-\alpha}$, where $S$ is the flux density, $\nu$ is the frequency and $\alpha=(\delta-1)/2$ is the particle spectral injection index. 
The spectral index is calculated as follows:
\begin{equation}
    \label{alpha}
    \rm{\alpha=-\frac{log(S_{\nu_1}/S_{\nu_2})}{log(\nu_1/\nu_2)}}
\end{equation}
and the error associated with the spectral index is:
\begin{equation}
     \rm{\alpha_{err}= \frac{1}{\ln \frac{\nu_1}{\nu_2}}\sqrt{\left(\frac{\delta \rm{S_1}}{\rm{S_1}}\right)^2+\left(\frac{\delta \rm{S_2}}{\rm{S_2}}\right)^2}}
 \end{equation}
 where $\rm{\nu_i}$ are the frequencies, $\rm{S_i}$ are the flux densities associated with the frequencies and $\rm{\delta S}$ is the absolute flux scale uncertainty.

The typical values of the spectral index for active radio galaxies are in the range 0.5-0.7 \citep{condon1992, giacintucci2012, giacintucci2021} and become higher with time, especially at high frequencies. On the other hand, the jets of the remnant radio galaxies such as NGC 6086 are switched off and the steepening should be higher, especially if the plasma has been ageing for a long time

\begin{figure}[!t]
    \centering
    \includegraphics[width=0.48\textwidth]{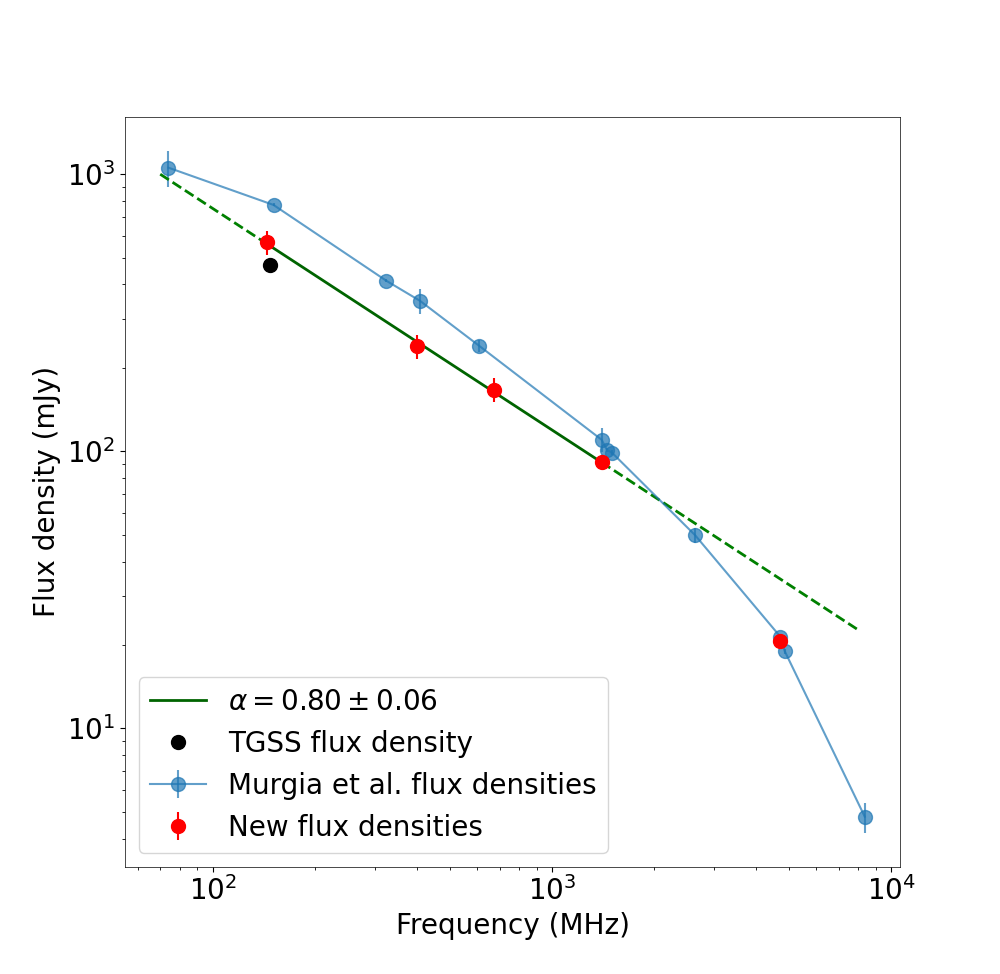}
    \caption{Integrated radio spectrum of the inner lobes of NGC 6086. Red circles show the flux density measurements presented in this paper, while blue circles show the flux densities listed in \cite{murgia2011} connected without fitting and the black point is the flux density measured by the TGSS all-sky survey of the GMRT at 147 MHz and 25 arcsec of resolution \citep{intema2017}. The green line represents the best linear fit using the values in the range 144-1400 MHz (with dashed extensions to emphasise the spectrum curvature and difference with respect to the literature flux density values). The best-fit spectral index is shown in the legend.}
\label{fig:integrated_spec}
\end{figure}

We measure the integrated radio spectrum in the inner lobes region and our result, presented in Fig. \ref{fig:integrated_spec}, is a power-law with a spectral index of 0.80$\pm$0.06 between 144-1400 MHz. At higher frequencies, the spectrum becomes steeper with a spectral index value of $\gtrsim$1.23 between 1400-4700 MHz. This implies a spectral curvature (SPC) defined as $\rm{\alpha_{high}-\alpha_{low}\sim0.4}$. We measure the integrated flux density at each frequency in the mid-resolution set of images, in a common region inside the 3$\sigma$ contours for all the frequencies involved and the values are reported in Table \ref{tab:flux} with the respective errors. The errors are dominated by the flux density scale errors, which we assume to be 10\% for both LOFAR \citep{shimwell2022} and uGMRT \citep{chandra2004} images and 5\% for the VLA \citep{perley2013} images.

We note that in the range 144-675 MHz we recover a lower flux density with respect to previous estimates reported in \cite{murgia2011}, as shown in Fig. \ref{fig:integrated_spec}. For the measurements at 151 MHz and 408 MHz (corresponding to the 7C survey, \citealp{riley1989}, the WENSS survey \citealp{rengelink1997} and the B2 survey \citealp{colla1970}), we suggest that the discrepancy is due to the larger beam used in these observations. To check the flux density at 610 MHz reported in \cite{murgia2011} obtained with the old GMRT we reprocessed the archival data with SPAM and found a flux density value consistent with our new observations. To further check the flux scale robustness we also checked the broad-band spectrum of the brightest compact sources in the field and did not recognise any clear offset. For these reasons, we rely on our flux density measurements for the rest of the analysis and do not apply any scale correction.

The total flux densities of the most significant morphological features are reported in Table \ref{tab:flux}. The flux densities of the filaments are measured from high-resolution images, in the mid-resolution set for the inner lobes and in the low-resolution set for the total flux densities. For the measures of the filaments, we use a common region above 3$\sigma$ for all the frequencies available at high resolution. The regions are highlighted in the LOFAR images shown in Appendix \ref{fig:substructures} both for the filaments and the inner lobes.

To generate spectral index maps we consider pixels above the 3$\sigma$ threshold at every frequency. The maps are presented in the following sections for the filaments, the inner lobes and the outer lobes.

\begin{table*}[!htp]
    \caption{Flux densities of NGC 6086. Measurements of the total flux densities  are performed on the images available for each substructure. The measurements of the filaments are performed on the high-resolution set; measurements of the inner lobes are performed on the mid-resolution set and measurements of the total emission are performed on the low-resolution set. Inner lobes and filaments measures are derived from a common region above the 3$\sigma$ threshold for all the frequencies of the same set.}
    \centering
    \begin{tabular}{c c c c c c}
    \hline\hline
        \multirow{2}{*}{Region} & S$\rm_{144 MHz}$ & S$\rm_{400 MHz}$ & S$\rm_{675 MHz}$ & S$\rm_{1.4 GHz}$ & S$\rm_{4.7 GHz}$\\
         & [mJy] & [mJy] & [mJy] & [mJy] & [mJy] \\ 
    \hline
    Inner lobes & 597$\ \pm\ $59  & 266$\ \pm\ $27 & 174$\ \pm\ $17 & 91.3$\ \pm\ $4.7 & 20.6$\ \pm\ $1.2\\
    Total & 878$\ \pm\ $88  & 276$\ \pm\ $28 & 203$\ \pm\ $20 & 91.3$\ \pm\ $4.7 & 20.6$\ \pm\ $1.2\\
    F1 & 7.81$\ \pm\ $0.86  & 2.30$\ \pm\ $0.26 & 1.36$\ \pm\ $0.15 &-&-\\
    F2 & 6.63$\ \pm\ $0.76 & 3.41$\ \pm\ $0.36 & 1.45$\ \pm\ $0.16 &-&-\\
    F3 & 4.78$\ \pm\ $0.54 & 1.22$\ \pm\ $0.14 & 0.78$\ \pm\ $0.09 &-&-\\
    \hline\hline  
    \end{tabular}
    \label{tab:flux}
\end{table*}

\subsection{Inner lobes}
\label{sec:inner_spec}
The inner lobes are the only substructures of the source that have been detected up to the highest frequency available in our analysis. For this reason, we exploit this wide frequency range (144-4700 MHz) with the mid-resolution set to analyse their emission in different parts of the synchrotron spectrum.

\begin{figure*}[!htp]
    \centering
    \minipage{0.48\textwidth}
        \centering
            \includegraphics[width=1\textwidth]{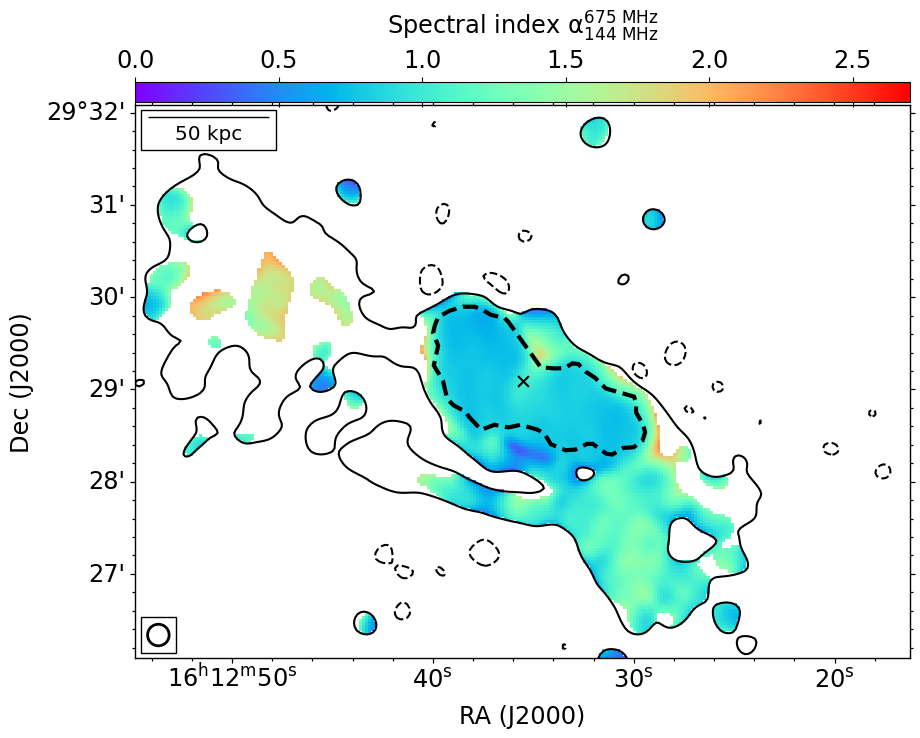}
        \endminipage\hfill
    \centering
    \minipage{0.48\textwidth}
        \centering
        \includegraphics[width=1\textwidth]{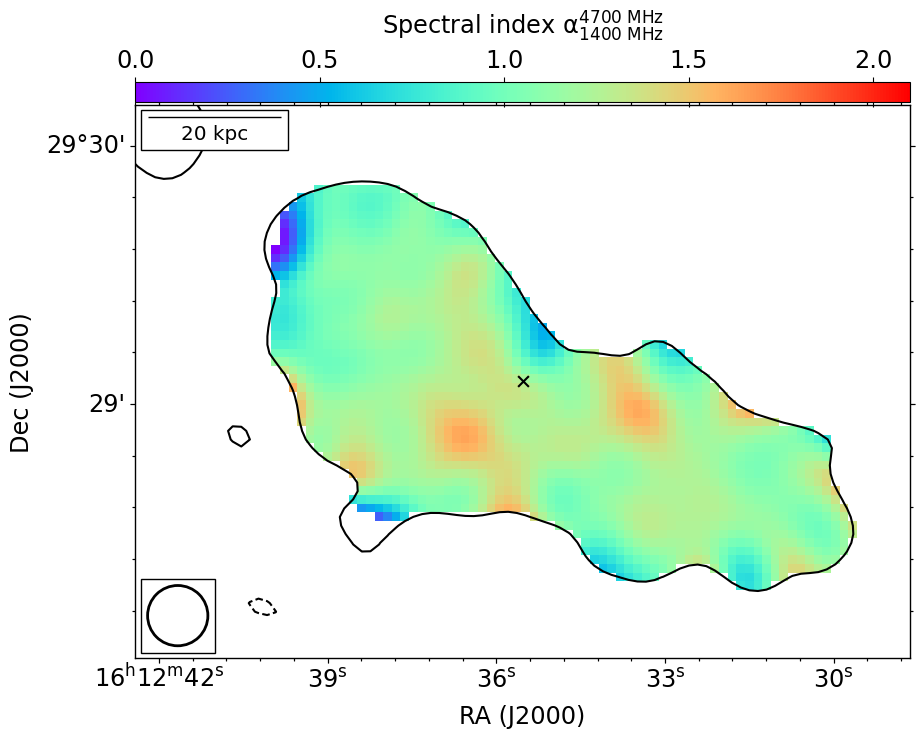}
    \endminipage\hfill
    \caption{NGC 6086 spectral index maps at a resolution of 14 arcsec, the beam is shown in the bottom-left corner, a reference scale is in the top-left one and the black `$\times$' marker is used to underline the location of the host galaxy. Left panel: spectral index map between 144-400 MHz overlaid with 3$\sigma$ LOFAR contour, the black dotted line region is the 4700 MHz 3$\sigma$ contour and is used as a reference for the inner lobes.
    Right panel: spectral index map between 1400 MHz and 4700 MHz with VLA 4700 MHz 3$\sigma$ contours overlaid.
    The associated error maps at low and high frequencies are shown in Appendix \ref{fig:6086_14_error_maps}.}
\label{fig:inner_spec_index}
\end{figure*}

Using the mid-resolution aligned radio images, we generate two spectral index maps that allow us to identify any spectral index gradient within the inner lobes. The first one is at low frequencies in the range 144-400 MHz and the second one is at high frequencies in the range 1400-4700 MHz, both shown in Fig. \ref{fig:inner_spec_index}. Source edges may have some artefacts that we exclude from our analysis and the error associated with those pixels are up to three times higher than the others (see Appendix \ref{fig:6086_14_error_maps} and \ref{fig:6086_30_error_map+filaments_error_map} for the of the error maps).

At both low and high frequencies, the spectral index map shows small variations throughout the entire inner lobes, the only visible trend is a mild steepening in the central region, which we further quantify below. The average spectral index value at low frequency is $\rm{\alpha^{400MHz}_{144MHz}\sim0.81}$ (similar to what is found in the radio integrated spectrum shown in Fig. \ref{fig:integrated_spec}) and at high frequency is $\rm{\alpha^{1400MHz}_{4700MHz}\sim 1.26}$, consistent with a curved spectrum due an ageing plasma.

We can already note that the spectral indices outside the inner lobes show steeper values but a more detailed analysis is made with the low-resolution images and presented in Sect. \ref{sec:outer_analysis}.


\begin{figure}[!htp]
    \centering
    \minipage{0.46\textwidth}
        \centering
            \includegraphics[width=1\textwidth]{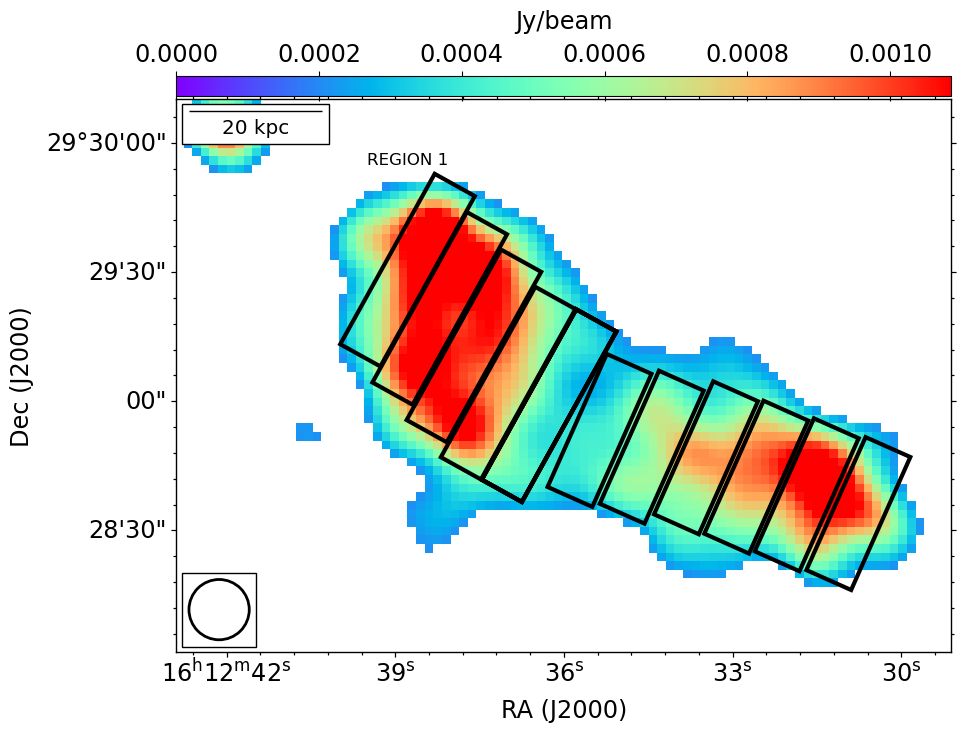}
        \endminipage\hfill
    \centering
    \minipage{0.46\textwidth}
        \centering
            \includegraphics[width=1\textwidth]{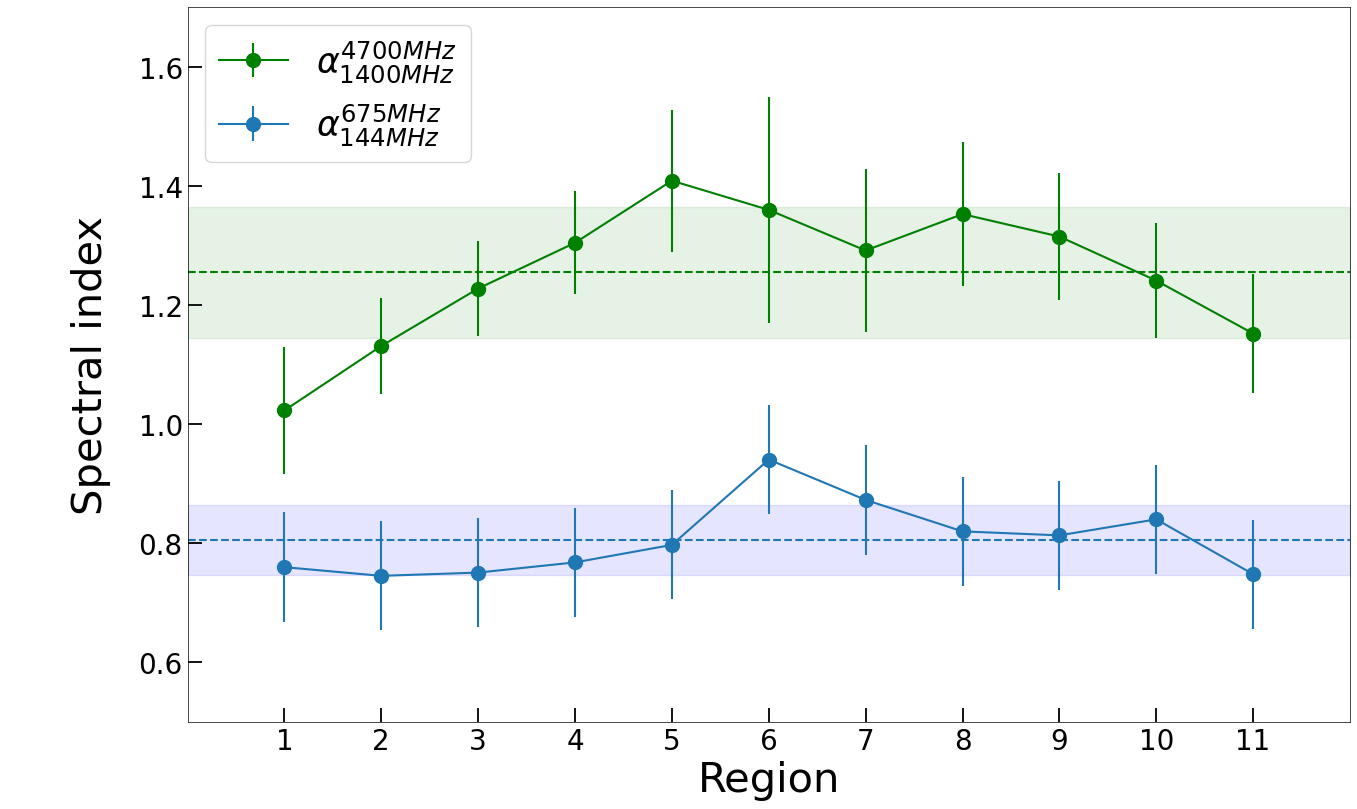}
        \endminipage
    \caption{Top: the image shows the inner lobes of the source at 14 arcsec resolution at 4700 MHz above the $3\sigma$ threshold. The eleven regions, larger than a beam, that are used for the spectral index analysis are overlaid. The beam is shown in the bottom-left corner and a reference scale is shown in the top-left one.
    Bottom: the plot shows the spectral index profile of the inner lobes in the two frequency ranges: 144-675 MHz and 1400-4700 MHz. The dashed lines represent the mean values while the coloured regions represent the standard deviations from the mean values.}
    \label{fig:inner_lobes_eleven_regions}
\end{figure}

To further explore the spectral trend across the inner lobes, we create spectral profiles using the regions shown in Fig. \ref{fig:inner_lobes_eleven_regions}. The regions are larger than the size of the beam and follow the most plausible direction of the jet (when they were active) as previously done by \cite{murgia2011}. We measure the flux density values inside the regions and derive the spectral index values using the formula presented in Sect. \ref{sec:spec}. The spectral indices are calculated between 144 MHz and 675 MHz and between 1.4 GHz and 4.7 GHz. In the plot in Fig. \ref{fig:inner_lobes_eleven_regions} (bottom panel) we show the spectral index profiles in both frequency ranges. The results obtained in the resolved spectral index maps are confirmed in the profile analysis:
\begin{itemize}
    \item the low-frequency spectral indices are systematically flatter than the high-frequency values in the same region. This is expected because, as seen in Fig. \ref{fig:integrated_spec}, the spectral break lies between 1.4 GHz and 4.7 GHz;
    \item the low-frequency spectral indices have uniform values with a mean of $\rm{\alpha^{675MHz}_{144MHz}\sim0.81 \pm0.06}$ with a mild steepening in the central regions;
    \item the high-frequency spectral indices have a wider range of values with a mean of $\rm{\alpha^{4700MHz}_{1400MHz}\sim1.26 \pm0.11}$ with a clear steepening, which is more evident than at lower frequencies and consistent with what previously found by \cite{murgia2011}.
\end{itemize}

\begin{figure}[!t]
    \centering
    \includegraphics[width=0.48\textwidth]{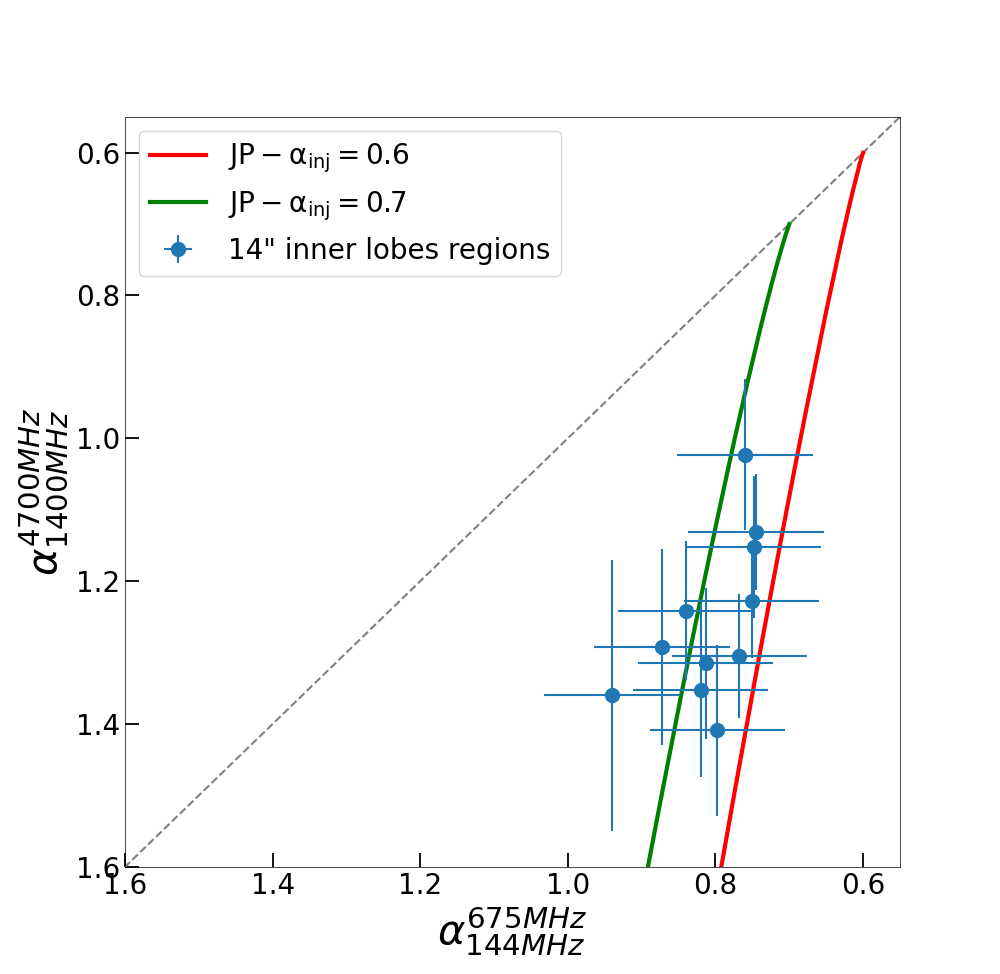}
    \caption{Colour-colour diagram of the inner lobes showing the spectral curvature of the plasma inside the selected regions. Two different spectral ageing model curves are overlaid to the point. The blue dots are the spectral index values that we measure in the eleven regions shown in Fig. \ref{fig:inner_lobes_eleven_regions}. The grey dashed line represents the bisector line where $\rm{\alpha^{400\ MHz}_{144\ MHz}}$ = $\rm{\alpha^{4700\ MHz}_{1400\ MHz}}$. The red line represents the JP single-injection evolution model with $\rm{\alpha_{inj}\sim0.6}$ while the green line is the same model with $\rm{\alpha_{inj}\sim0.7}$.}
\label{fig:cc}
\end{figure}

To visualise the spectral curvature and have an immediate comparison with radiative models, we have used the colour-colour plot \citep{katzstone1993, shulevski2017, brienza2020} (see Fig. \ref{fig:cc}). In this plot, a pair of spectral indices are plotted on the x and y axis respectively. In our case we plot $\alpha^{675\ \rm{MHz}}_{144\ \rm{MHz}}$ in the x-axis and $\alpha^{4700\ \rm{MHz}}_{1400\ \rm{MHz}}$ in the y-axis. To create the diagram we use the eleven regions shown in Fig \ref{fig:inner_lobes_eleven_regions}. 

The observed values are all below the bisector line ($\rm{\alpha^{675\ \rm{MHz}}_{144\ \rm{MHz}}=\alpha^{4700\ \rm{MHz}}_{1400\ \rm{MHz}}}$, shown as a dashed line), which implies steepening at the higher frequencies for all the regions, consistent with particle ageing. In the plot, two simulated tracks of spectral evolution are drawn as a reference and they are representative of two Jaffe-Perola (JP) \citep{jaffe1973}, single-injection, models. We choose the JP model because it is suited for populations that have been accelerated at the same time. The two models have a different injection index (spectral index at the moment of the injection, where the model intersects the bisector line), in particular, the red line has the value of the injection index found by the software (see Sect. \ref{sec:duty_cycle} for details) for the inner lobes which is $\rm{\alpha_{inj}\sim0.6}$. The particles are ageing passively since $\sim20$ Myr and for this reason a higher injection index ($\rm{\alpha_{inj}\sim0.7}$, green line) seems to be more representative for some of the points.

\subsection{Outer lobes}
\label{sec:outer_analysis}
Here, we extend the spectral analysis of NGC 6086 to the outer lobes. As presented in the right column of Fig. \ref{fig:6086}, this large-scale diffuse emission is only visible in low-resolution images.

\begin{figure}[!h]
    \centering
    \minipage{\columnwidth}
        \centering
            \includegraphics[width=1\textwidth]{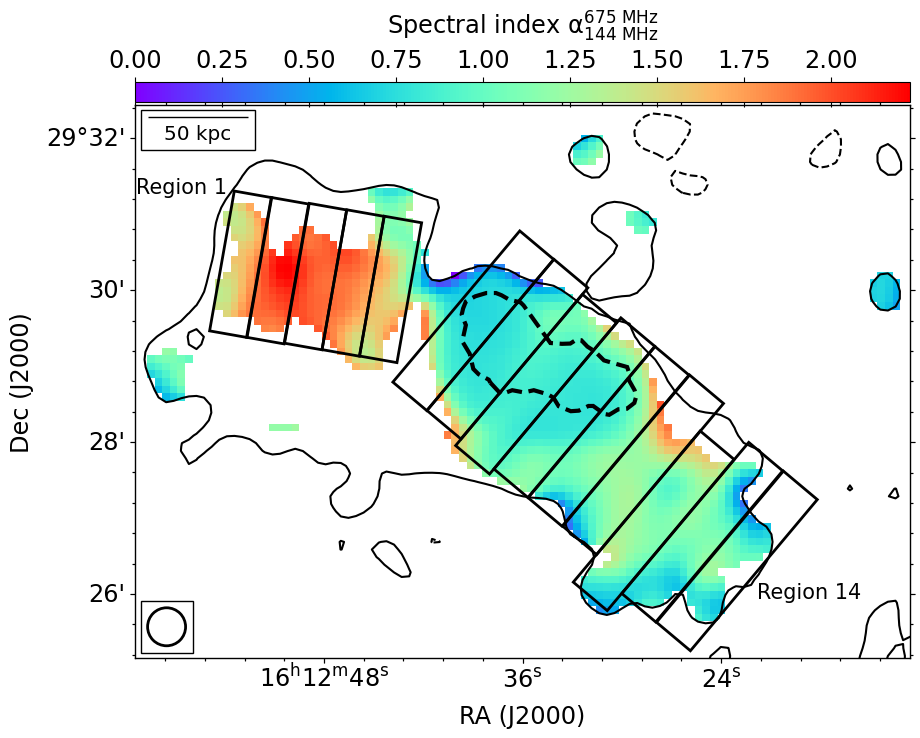}
        \endminipage\hfill
    \minipage{\columnwidth}
        \centering
            \includegraphics[width=1\textwidth]{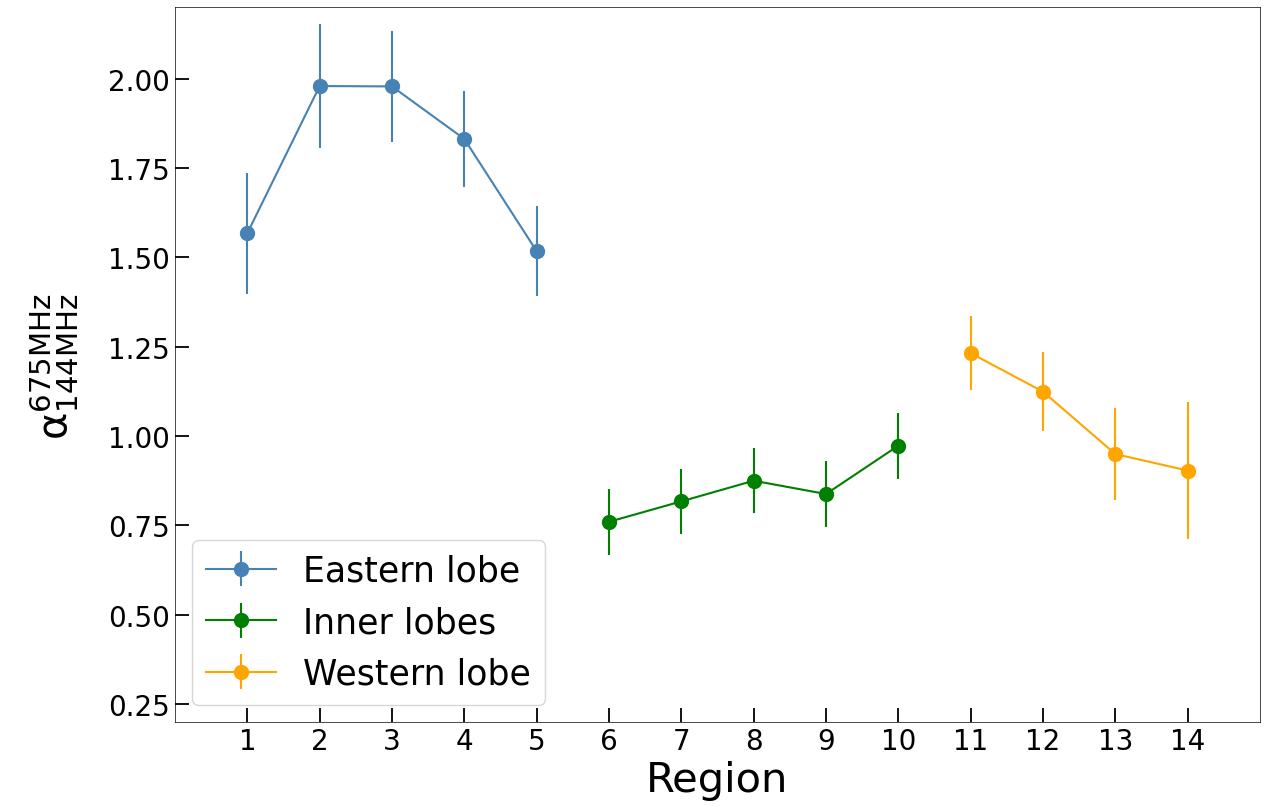}
        \endminipage
    \caption{Top: NGC 6086 spectral index map between 144 MHz and 675 MHz at low-resolution (30 arcsec). Contours show the LOFAR emission and are drawn at (-3, 3)$\times\sigma$. The black dotted line is the inner lobes region, defined by the 3$\sigma$ contour of the VLA 4700 MHz and used as a reference. Fourteen regions across the entire source are selected in the map to look for the spectral index trends. The beam is shown in the bottom-left corner and a reference scale in the top-left one. The error map is presented in Appendix \ref{fig:6086_30_error_map+filaments_error_map}. Bottom: the plot shows the spectral index values between of the fourteen regions. We colour-code the three main regions of the large-scale diffuse emission detected in the low-resolution images.}
\label{fig:30_14_regions}
\end{figure}

Therefore, to perform the resolved analysis of the outer lobes, we use the low-resolution set of images. We create a spectral index map using the 144 MHz and the 675 MHz images. At 400 MHz the extended emission is not entirely recovered due to sensitivity limits so we do not include it in the spectral index map, shown in Fig. \ref{fig:30_14_regions}. 

The spectral index values in the outer lobes are steeper than the inner ones as expected for an ageing plasma. It is worth noticing that the two outer lobes show quite different spectral index values. The spectral index of the eastern lobe reaches values up to $\rm{\alpha^{675MHz}_{144MHz}\sim2.12 \pm0.20}$, while in the western lobe up to $\rm{\alpha^{675MHz}_{144MHz}\sim1.25 \pm0.12}$.

We also derive a lower limit of the spectral index of the eastern lobe in a region detected only in the LOFAR image at 3$\sigma$. We find a value equal to $\rm{\alpha_{144\ MHz}^{675\ MHz}\geq1.6}$, which is lower than the value measured in the rest of the lobe, suggesting that the non-detection is mainly due to a sensitivity limit rather than to a steeper spectral index.

To further investigate the spectral properties of the outer lobe, we create a spectral index profile using low-resolution images, similarly to what done for the inner lobes. We create fourteen rectangular regions, with different sizes and with area larger than one beam (see Fig. \ref{fig:30_14_regions}, top panel, for the selected regions). We have plotted the $\alpha^{675\rm{MHz}}_{144\rm{MHz}}$ values in Fig. \ref{fig:30_14_regions}, bottom panel, and we can conclude that:
\begin{itemize}
    \item we confirm the steeper values of the spectral index in the eastern lobe with respect to the other one;
    \item there are different trends in the spectral index values by moving towards the host galaxy in the two outer lobes. The eastern lobe shows a spectral flattening moving both towards the host galaxy and the outer regions. On the contrary, the western lobe shows a clear steepening towards the host (similar to what is observed in the inner lobes).
\end{itemize}

\subsection{Filaments}
\label{sec:filaments}

\begin{figure}[!htp]
    \centering
    \includegraphics[width=0.48\textwidth]{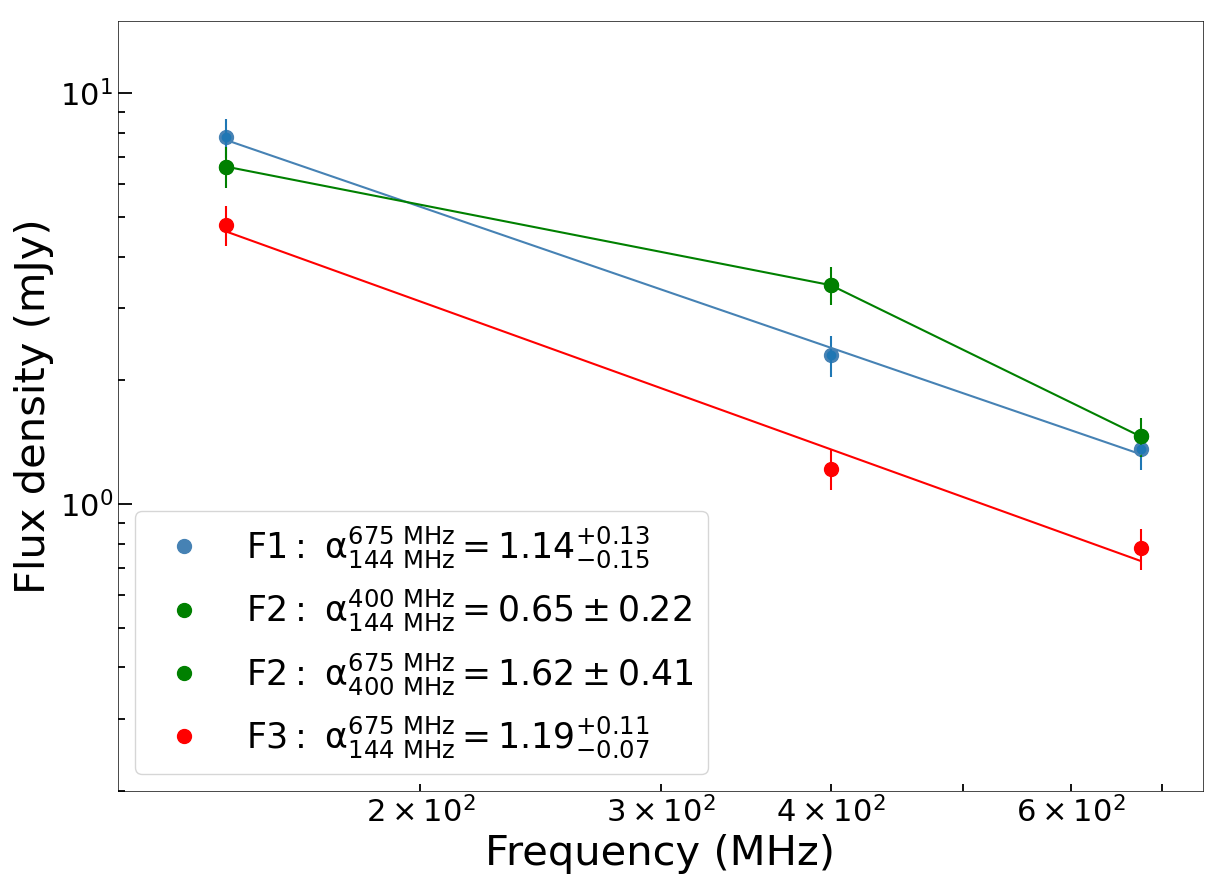}
    \caption{The image shows the spectra of the three filaments detected in the high-resolution images. We have added two best-fit power-laws for filaments F1 and F3 while the spectrum of filament F2 is not consistent with a power-law due to the spectral steepening. For this reason, we have measured the spectral index by linking the three flux density values. The legend is shown in the bottom-left corner along with the spectral index value.}
    \label{fig:filaments_fit}
\end{figure}

To perform the spectral analysis of the filaments, we use the high-resolution set of images including all three frequencies below 1 GHz where they are detected. We have measured the flux density inside the filaments in a common region above 3$\sigma$ for all three frequencies (the regions are highlighted in the LOFAR image in Appendix \ref{fig:substructures}). The values are listed in Table \ref{tab:flux}. The integrated spectra of the filaments are shown in Fig. \ref{fig:filaments_fit}. 

From this, we can see that the filaments F1 and F3 have a spectrum consistent with a power-law between 144 MHz and 675 MHz with spectral index 1.14$^{+0.13}_{-0.15}$ and 1.19$^{+0.11}_{-0.07}$, respectively. Instead, F2 shows a spectral break and its spectral indices are $\rm{\alpha^{400\ MHz}_{144\ MHz}=0.65\pm0.22}$ before the break and $\rm{\alpha^{675\ MHz}_{400\ MHz}=1.62\pm0.41}$ after the break.

\begin{figure}[!h]
    \centering
    \minipage{\columnwidth}
        \centering
            \includegraphics[width=1\textwidth]{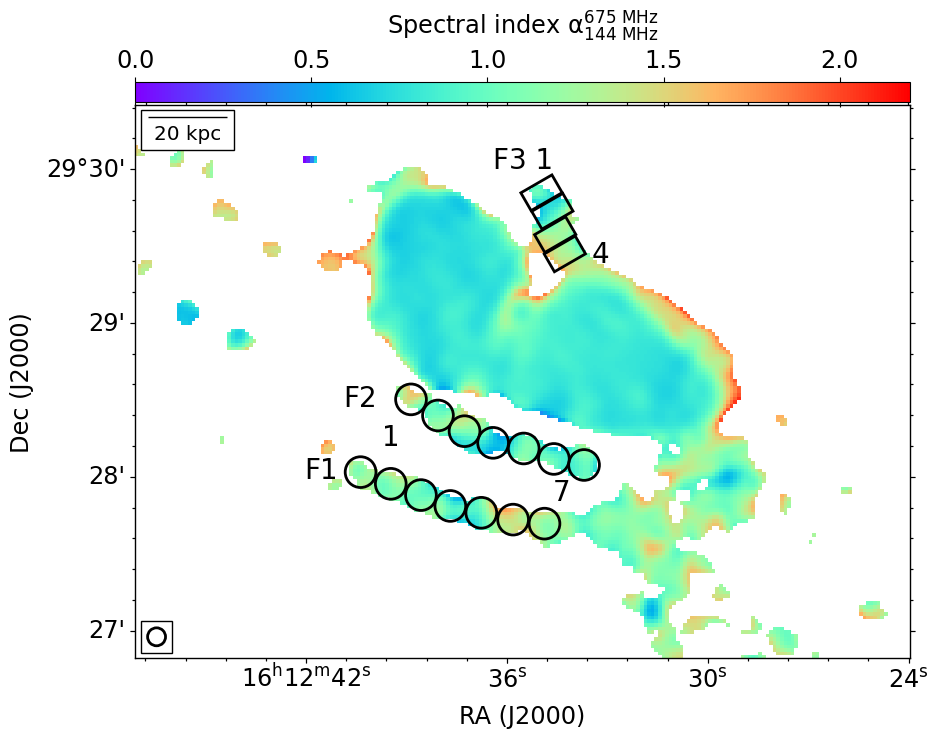}
        \endminipage\hfill
    \minipage{\columnwidth}
        \centering
            \includegraphics[width=1\textwidth]{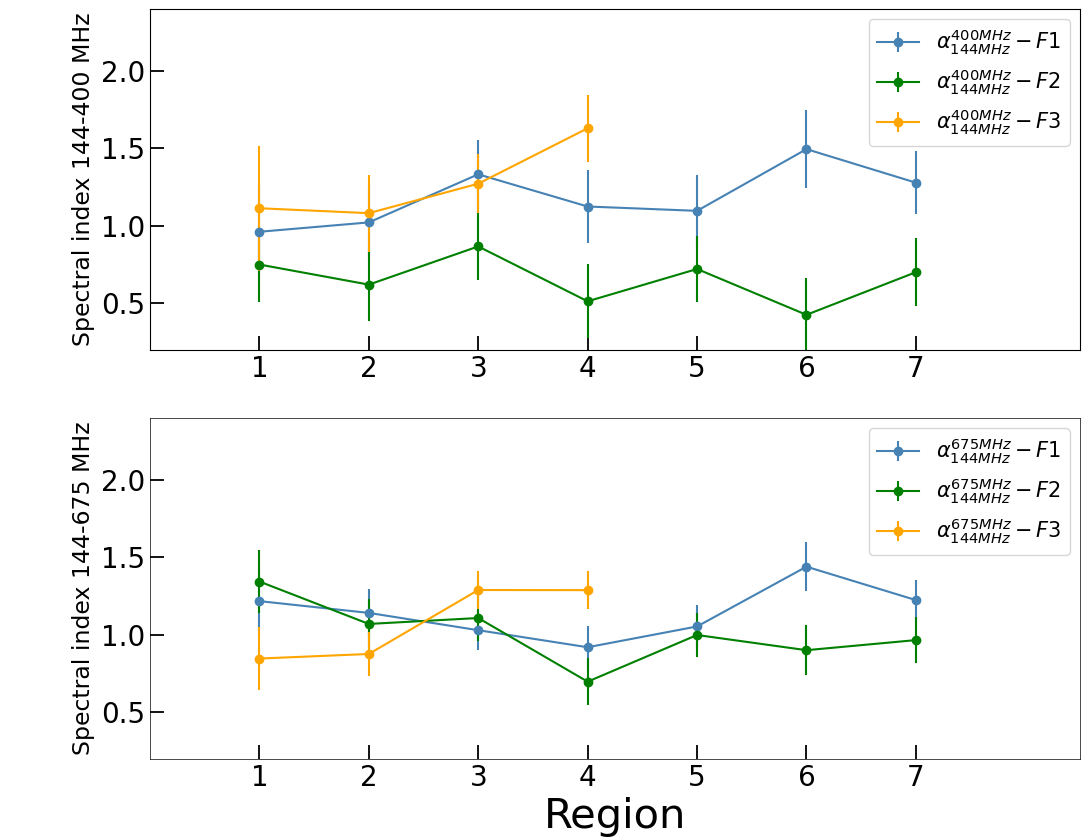}
        \endminipage
    \caption{Top: spectral index map of NGC 6086 at a resolution of 7 arcsec. Only pixels above 3$\sigma$ both for LOFAR 144 MHz and uGMRT 675 MHz images are considered. Eighteen regions across the three filaments are selected in the map to investigate spectral index trends. The beam is shown in the bottom-left corner and a reference scale is in the top-left one (the error map is presented in Appendix \ref{fig:6086_30_error_map+filaments_error_map}). Centre and bottom: the plots show the spectral index trends in the filaments colour-coded as shown in the legend. The central panel refers to $\rm{\alpha_{144MHz}^{400MHz}}$ while the bottom panel refers to $\rm{\alpha_{144MHz}^{675MHz}}$}
\label{fig:7_image_and_spix}
\end{figure}

To further analyse the three filaments, we create a spectral index map using the 144 MHz and 675 MHz images and measured the spectral index across the structures looking for possible trends. We choose seven circular regions larger than a beam for filaments F1 and F2 and four rectangular regions for F3. We report in Fig. \ref{fig:7_image_and_spix} the spectral index map between 144-675 MHz at 7-arcsec resolution showing the eighteen selected regions in the top panel. From the spectral index profiles shown in the bottom panels of Fig. \ref{fig:7_image_and_spix} we can see that the spectral index does not show any clear trend along the filaments.

\section{Spectral age analysis}
\label{sec:age}
In this section, we present a spectral age study of NGC 6086. The analyses for the inner and outer lobes are performed following different approaches, as a result of the various image sets available for each structure. For both pairs of lobes, we use spectral evolution models to give first-order constraints to the age of the plasma. To do this, we have used the Broadband Radio Astronomy Tools\footnote{\url{http://www.askanastronomer.co.uk/brats/}} (\texttt{BRATS}) \citep{harwood2013, harwood2015} software, which allows for detailed spectral analysis of radio images, providing a suite of model fitting, visualisation and statistical tools. The evolution of the synchrotron spectrum is determined by the radiative losses that particles have undergone and is directly related to the age of the plasma. The synchrotron ageing depends on the magnetic field, the redshift and the break frequency. The classical equation of the radiative age is defined as:
\begin{equation}
\label{formula:age}
    \rm{t_{age} \sim 3.2 \times 10^{10}\frac{B^{0.5}}{B^2+B^2_{CMB}} [(1+z)\nu_{break}]^{-0.5}\ \rm{yr}}
\end{equation}
where $\rm{B\ [\mu G]}$ is the magnetic field intensity of the source, $\rm{B_{CMB}}$ is the magnetic field intensity required to have a synchrotron emission that is equivalent to the energy losses via IC scattering and its value is $\rm{B_{CMB}=3.25\ (1+z)^2\ [\mu G}]$, $z$ is the redshift and $\rm{\nu_{break}\ [MHz]}$ is the frequency. The formula shows that the higher the emission break frequency and the higher the magnetic field intensity, the lower the lifetime. The radiative age formula gives a first-order age value, but a more accurate analysis can be made by using the integration of radiative cooling equations as done by the \texttt{BRATS} software.

\subsection{Inner lobes}
The inner lobes are detected at all five frequencies, for this reason, we generate a spectral age map by performing a pixel-to-pixel spectral model fitting using the mid-resolution (14 arcsec) images set. We fit the JP model to the flux density data of the five images. This model assumes that all the relativistic particles are accelerated by the same event at the same time. We apply this model to all the single pixels of the image, assuming that they are accelerated at the same time. The model we choose requires the setting of two parameters for computing the age: the spectral injection index and the magnetic field value.

The injection index corresponds to the spectral index at the moment of injection. Using BRATS, we derive the best injection value by fitting the data with different injection indices. For the inner lobes of NGC 6086, we find that the $\chi^2$ is minimised for$\rm{\alpha_{inj}=0.60}$, in good agreement with the value of the majority of the radio-loud AGN \citep{webster2021}, which is in the range 0.5-0.7, and consistent with those found in the B2 sample of radio galaxies \citep{colla1970}.

To compute the magnetic field intensity, we assume equipartition conditions and use the formula presented by \cite{murgia2012} as reference:
\begin{equation}
\label{formula:b_eq_murgia}
     \rm{B_{eq}=\left[\frac{4\pi m_e c^2 (1+\alpha)(1+k) L_\nu (\gamma_{max}^{1-2\alpha}-\gamma_{min}^{1-2\alpha})}{\nu^{-\alpha}(C_\alpha)_{LOS}(1-2\alpha)V}\right]^{(3+\alpha)^{-1}} = 4.4\ \mu \rm{G}}
\end{equation}
where $\gamma$ values are the Lorentz factors range of the electron population, assumed as $\rm{\gamma_{max}=10^6}$ and $\rm{\gamma_{min}=10^2}$; $\rm{m_e}$ is the electron mass; $\rm{\alpha=0.6}$ is the injection index found with the model fitting in \texttt{BRATS}; $\rm{k=0}$ is used assuming no contribution to the energy density by the proton particles and redshift $z=0.0318$. The volume of the inner lobes is modelled as two ellipsoids with the length of the third axis equal to the shorter one. In particular, the eastern lobe is 24 $\times$ 16 $\times$ 16 kpc$^3$ and the western one is 26 $\times$ 14 $\times$ 14 kpc$^3$.

As a reference, we also consider a more conservative magnetic field value:
\begin{equation}
    \rm{B_{min}=\frac{B_{CMB}}{\sqrt{3}}=\frac{3.25(1+z)^2}{\sqrt{3}}= 2.0\ \mu \rm{G}}
\end{equation} 
which corresponds to the minimum possible radiative losses of the synchrotron emission for the plasma at a given redshift. With this magnetic field, the spectrum of the plasma will steepen more slowly than with the equipartition value. The minimum magnetic field is a lower limit and it is useful to provide an upper limit to the spectral age values.

\begin{figure}[!htp]
    \includegraphics[width=0.46\textwidth]{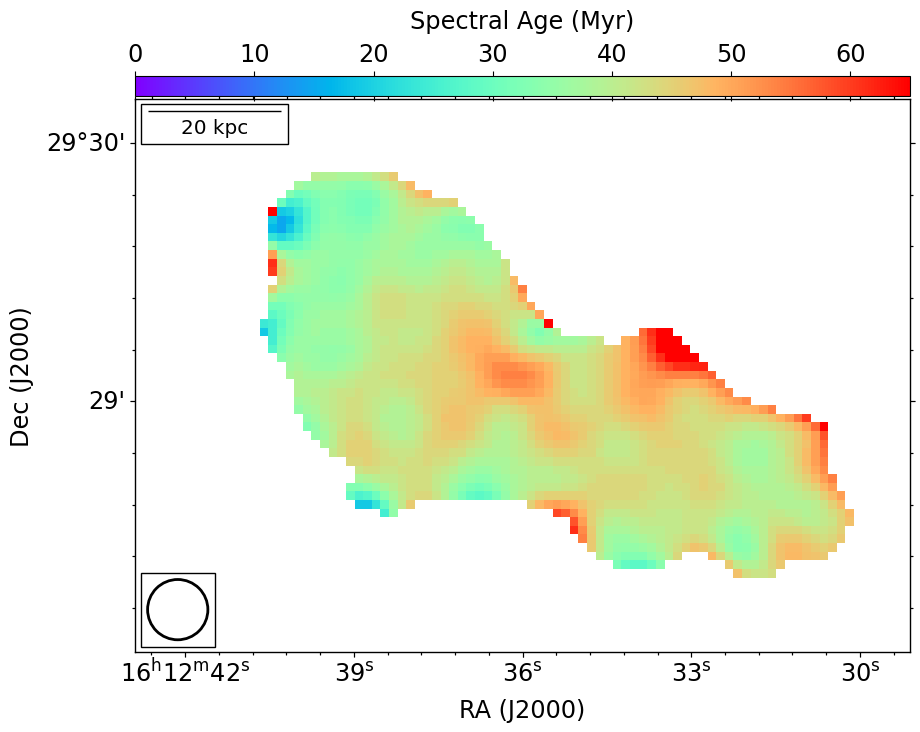}
    \caption{NGC 6086 spectral age map, obtained using the five mid-resolution images and assuming a JP model and equipartition magnetic field value. The beam is shown in the bottom-left corner and a reference scale is in the top-left one. The error map is presented in Appendix \ref{fig:age_error_map}.}
\label{fig:spec_age_map}
\end{figure}

The final spectral age image of the inner lobes using $\rm{B_{eq}}$ is shown in Fig. \ref{fig:spec_age_map}. It is important to stress that the change in magnetic field intensity affects the age of the plasma and range of values, but does not affect the relative age trend across the source. 

We can see in the age map that the oldest populations are located in the central part of the lobes, near the position of the host galaxy. The results are summarized in Table \ref{tab:ages_inner}. We define $\rm{t_{max}}$ and $\rm{t_{min}}$ as the oldest and the youngest age value observed in the map over the entire source. The error values are derived from the error map shown in Appendix \ref{fig:age_error_map}.

\begin{table}[!htp]
    \centering
    \caption{The table shows the results of the spectral age analysis of the inner lobes. We report the values associated with the last active phase event for both magnetic field assumptions.}
    \large
    \begin{tabular}{l|cc}
         \hline\hline
         & $\rm{t_{max}}$ & $\rm{t_{min}}$\\
         & [Myr] & [Myr]\\
         \hline
         \multirow{2}{*}{$\rm{B_{eq}}$} &  \multirow{2}{*}{51 $\pm$ 5} & \multirow{2}{*}{33 $\pm$ 2}\\\\
         \hline
         \multirow{2}{*}{$\rm{B_{min}}$} &  \multirow{2}{*}{69 $\pm$ 7} & \multirow{2}{*}{45 $\pm$ 4}\\\\
         \hline\hline
    \end{tabular}
    \label{tab:ages_inner}
\end{table}

\subsection{Outer lobes}
\label{outer_age}
The outer lobes are only recovered at the three lowest frequencies and, for this reason, we cannot generate a spectral age map of the outer lobes as we did with the inner ones. Instead of an age map, we make a first-order age estimate by generating a series of simulated JP models with \texttt{BRATS} at different time steps (1 Myr). For each of these simulated models, we derive a spectral index value in the frequency range 144-675 MHz and compare it with the value measured in our observations. In particular, we measured in both the outer lobes the highest and the lowest mean value of the spectral index in a beam-size region using the spectral index map at low resolution (see Fig. \ref{fig:30_14_regions}, top panel). The model whose spectral index matches the observations is used to infer the age of the plasma. We use the same method to measure the value of the spectral index errors in the same regions and derive an age range.


To create the simulated JP spectra, we use the same parameters as for the inner lobes (injection index and magnetic field values). It is plausible to think that the equipartition magnetic field intensity is lower in the outer lobe than in the inner ones due to the expansion of the plasma. For this reason, the values calculated with the same magnetic field should be treated as lower limits to the plasma age. The results are listed in Table \ref{tab:ages_outer}. 

\begin{table}[!htp]
    \centering
    \caption{The table shows the results of the spectral age analysis of the outer lobes for both magnetic field assumptions. The age differences between the two lobes reflect the spectral index differences seen in Sect. \ref{sec:outer_analysis}.}
    \begin{tabular}{l|cccc}
         \hline\hline
         & \multirow{2}{*}{Lobe} & $\rm{t_{max}}$ & $\rm{t_{min}}$ & $\rm{t_{on}}$\\
         & &[Myr] & [Myr] & [Myr]\\
         \hline
         \multirow{2}{*}{$\rm{B_{eq}}$} & E &  202 $\pm$ 15 & 142 $\pm$ 17 & 60 $\pm$ 32 \\
                & W &  122 $\pm$ 14 &  92 $\pm$ 18 & 30 $\pm$ 32 \\
         \hline
         \multirow{2}{*}{$\rm{B_{min}}$} & E &  274 $\pm$ 21 & 192 $\pm$ 24 & 82 $\pm$ 43 \\ 
                & W &  166 $\pm$ 18 & 125 $\pm$ 25 & 41 $\pm$ 43  \\
         \hline\hline
    \end{tabular}
    \label{tab:ages_outer}
\end{table}

\section{Duty cycle constraints}
\label{sec:duty_cycle}
In this section, we use the spectral age derived in the previous section to investigate the duty cycle of the radio galaxy associated with NGC 6086. We consider the inner lobes and the outer lobes as inflated by two different active phases of the AGN jets. We summarise in Tables \ref{tab:ages_inner} and \ref{tab:ages_outer} the estimates of the age of the different structures. The ages of the outer lobes are different, for this reason, we decide to consider the mathematical average between their values as a reference for the estimation of the duty cycle.

We measure the active periods by computing the difference between the age of the oldest and the youngest observed particle population inside a beam-size region. Furthermore, we use the average value of the age found in the two outer lobes. The first observable phase of the jet activity (which gave rise to the outer lobes) lasted $\rm{t_{on}\ (B_{eq})=t_{max}-t_{min}=45\ Myr}$ with the equipartition magnetic field and lasted $\rm{t_{on}\ (B_{min})=61\ Myr}$ with the minimum value. We note that these values refer to the age range of the diffuse region detected by both the LOFAR and uGMRT band-4 observation, but the LOFAR detection shows a more vast area that could not be investigated at the moment.

To estimate the age of the latest observable phase of jet activity (which gave rise to the inner lobes), we measure the maximum and minimum age values in the spectral age map inside a beam-size region. The oldest populations are found in the central part of the inner lobes, where the host galaxy is located, while the youngest populations are located at the edges of the lobes. The resulting active time periods of the inner lobes are $\rm{t_{on}\ (B_{eq})=t_{max}-t_{min}=18\ Myr}$ and $\rm{t_{on}\ (B_{min})=24\ Myr}$.

All the jets activity timescales are summarised in Table \ref{tab:duty_cycle}. Once again, we highlight that the following values do not take into account any dynamical evolution of the plasma (i.e. compression, expansion, magnetic field variation, etc.).

\begin{table*}[!ht]
    \centering
    \caption{Estimates of the duty cycle timescales using spectral age results, obtained using the JP model on the inner and outer lobes. The duty cycle is presented for the two different values of the magnetic field intensity discussed in the text. The ages are expressed in Myr ago and $\rm{\Delta t}$ represents the duration of the phase in Myr.}
    \begin{tabular}{c|cc|cc}
        \hline\hline 
        \multirow{2}{*}{Phase} &  \multicolumn{2}{c}{$\rm{B_{eq}}$} & \multicolumn{2}{|c}{$\rm{B_{min}}$}\\
        & [Myr (ago)] & $\Delta$t [Myr] & [Myr (ago)] & $\Delta$t [Myr]\\
        \hline
        First active & $162\pm14 - 117\pm17$ & $45\pm31$ 
        & $220\pm19-159\pm24$ & $61\pm43$\\
        First inactive & $117\pm17-51\pm5$ & $66\pm24$ 
        & $159\pm24-69\pm7$ & $90\pm26$ \\
        Second active & $51\pm5-33\pm2$ & $18\pm7$ 
        & $69\pm7-45\pm4$ & $24\pm13$\\
        Second inactive & $33\pm2-0$ & $33\pm2$ 
        & $45\pm4-0$ & $45\pm4$\\
    \hline\hline
    \end{tabular}
\label{tab:duty_cycle}
\end{table*}

The total age of the plasma inflated by the radio jets has a first-order duration of $\sim$162 Myr under the equipartition assumption and becomes $\sim$1.35 times longer under the minimum magnetic field assumption. We can consider the latter value as a lower limit for the magnetic field implying an upper limit for the age results. In both cases, the fraction of time the source is active is $\sim$39\%, measured as $\rm{t_{on}/(t_{on}+t_{off})}$, where $\rm{t_{on}}$ is the sum of the two active phases and $\rm{t_{off}}$ is the sum of the two inactive ones.\\

We remind here that the aforementioned values are subject to a number of uncertainties, which we briefly comment on here. 

The magnetic field is the major source of uncertainty in the duty cycle estimate because it is a fundamental value that influences all the age measurements and the evolution of the spectral shape. We cannot measure it directly, for this reason, we use the formula of the equipartition value (see Eq. \ref{formula:b_eq_murgia}). In the estimate of the magnetic field intensity are included a few factors ($\rm{\gamma_{max}}$ and $\rm{\gamma_{min}}$, the volume of the lobes and the k value) which are poorly known and above all, even the equipartition assumption is not necessarily correct. In particular, according to \cite{croston2018}, the contribution of cosmic ray protons is different between FRI and FRII, and the equation could have $\rm{k\neq0}$ as a consequence.
We assume the same magnetic field for both the inner and outer lobes. It is reasonable to think that, in the outer lobes, the magnetic field has a lower value than in the inner ones due to the expansion of the plasma. However, we decide not to compute another value of the equipartition magnetic field to avoid the introduction of new uncertainties related to the measurement of the parameters. Despite all this, we are confident that the true value lies in the range $\rm{B_{eq} - B_{min}}$ and this dictates the main uncertainties on the age values.

Another source of uncertainties comes from the assumption that the plasma is passively ageing and is not subject to any dynamic process, such as expansion or compression. These phenomena indeed can alter the spectral shape and, consequently, the age estimate \citep[e.g.][]{biava2021}. In the case of NGC 6086, it is reasonable to assume that at least the outer lobes may have suffered some adiabatic expansion and not necessarily uniformly for the eastern and western lobes (see Sect. \ref{sec:discussion} for more details on this). The effect of this would be to increase the age of the plasma with respect to the simple ageing model. 

The duty cycle results imply that a magnetic field change by a factor greater than two affects the ages and increases them by a factor of $\sim1.35$. Despite different magnetic field intensities, the total active time remains the $\sim39\%$ since the first of the last two jet outbursts. Although all the possible sources of uncertainties for the age, we are confident about the duty cycle conclusions and even if the equipartition value is the result of a few assumptions, it is reasonable to think that the actual ages of the active and inactive phases are located inside the age ranges derived from the two magnetic field values used above.\\

\section{Discussion}
\label{sec:discussion}
In this work, we presented a broadband radio frequency analysis of the radio source associated with the galaxy NGC 6086, the central galaxy of the Abell 2162 group.
We used three new deep low-frequency observations: LOFAR HBA at 144 MHz, uGMRT band-3 at 400 MHz and uGMRT band-4 at 675 MHz. We complemented these with two archival VLA observations at higher frequencies (1.4 GHz and 4.7 GHz). Beyond the already-known double lobes, our new low-frequency images have revealed previously undetected diffuse emission surrounding the source, which we interpret as associated with a past active phase of the AGN of the host galaxy. We refer to these structures as inner and outer lobes, respectively. Within this new emission, we also detect three interesting filamentary structures, whose possible origin we discuss below.



As already mentioned in Sect. \ref{sec:overview}, previous analyses of the inner lobes have classified them as a remnant radio galaxy \citep{parma1986, owen1997, giacintucci2007,liuzzo2010,murgia2011}, that is a radio galaxy whose jets and nuclear engine have switched off. Thanks to the detection of the new outer lobes we can now classify the source as a restarted radio galaxy, defined as a galaxy in which we can see multiple phases of jet activity at the same time. In particular, the larger extension of the outer lobes compared to the inner ones suggests that the expansion in the environment could be the reason for this difference, but the higher power of the jets in the previous event could also have played an important role.

In the literature, there are several cases observed of multiple AGN jet activity (e.g. \citealp{schoenmakers2000b, schoenmakers2000, fabian2005, konar2006, wise2007, nandi2012, konar2013, brienza2020, giacintucci2021, biava2021}), with only very few of them in galaxy groups environment (e.g. \citealp{randall2015, maccagni2020, schellenberger2021, Schellenberger2023}). What makes NGC 6086 unique is that it shows two generations of remnant lobes, while in most other cases we see a pair of active lobes and a pair of remnant lobes. This is especially true if we consider the galaxy group environment, for example, NGC 507 is one of the rare sources such as NGC 6086 with two pairs of remnant lobes \citep{brienza2022}. 

The outer lobes of NGC 6086 are slightly bent in the south-east direction, probably because of the motion of the galaxy within the group, and they are not clearly sharply detached as is the case in most double-double galaxies. Restarted radio galaxies without a double-double morphology, however, are not uncommon (e.g. \citealp{jamrozy2007, jurlin2020, kukreti2022}.)
The reason could be the low power of the jets and the consequent inflation of plasma in the surroundings of the host galaxy or the confinement of the plasma in the IGrM. 
Thanks to the unique detection of two pairs of remnant lobes, we have the rare opportunity to probe the full duration of two full phases of jet activity and thus get a clear estimate of the duty cycle of an AGN at the centre of a galaxy group. This is a very important piece of information to quantify the energetic impact that the AGN has on its surrounding environment.


The average spectral index value that we measure across the inner lobes is $\sim 0.81$ and $\sim 1.26$ at low and high frequencies, respectively, consistent with plasma ageing. The spectral index distribution of the inner lobes at low frequencies (144-675 MHz) is presented in this work for the first time and shows similar trends to the one observed at high frequencies (1400-4700 MHz), already reported by \cite{murgia2011}. It is steeper in the vicinity of the centre of the inner lobes and flatter towards the edges. This may suggest that, when the jets were active, particle acceleration was mostly happening at the edge of the source rather than close to the core, as observed in FRII radio galaxies \citep{harwood2017} or lobed FRI \citep{laing2008}.
 
However, the observed spectral trend is very mild. Projection effects and the resulting superposition of different particle populations could play an important role in smoothing out any spectral index trends. On top of that, a physical explanation is given by \cite{murgia2011}, who suggested that remnant radio sources for which the duration of the quiescent phase is a significant fraction of the lifetime, should all be characterised by a uniform spectral index distribution along the fading lobes. These authors showed that after the jet quenching any pre-existing spectral index gradient along the lobes is rapidly erased as the breakup frequency reaches approximately the same value in every part of the source. As the age of the off-phase progresses, the spectral index of the lobes systematically increases but with smaller and smaller variations from point to point.

Thanks to the increased spectral coverage with respect to previous works, we derived the spectral age of the inner lobes by performing a resolved analysis using all the five frequencies available (see Table \ref{tab:ages_inner}). The method used to estimate the age of the radio plasma is presented in other works and in different sources and environments \citep[e.g.][]{harwood2013, shulevski2017, brienza2020, rao2023}. We measure two active phases that last $\sim$45 Myr and $\sim$18 Myr, using the equipartition magnetic field assumption. The two inactive periods last $\sim$66 Myr and $\gtrsim$33 Myr, resulting in a total active time of $\lesssim$39\%. 
We remark that our knowledge of the duty cycle is limited to the last two active phases. How fast the duty cycle was in earlier epochs cannot be constrained due to the absence of visible radio emission nor we can assume how it will be in the future.

The age of the inner lobes is consistent with what was previously found by \cite{murgia2011} by fitting the integrated spectrum with a CI OFF model. From the fit of the integrated spectrum, they found a comparable time of 55 Myr and consistent active and inactive period ages. From the fit of the spectral index trend, they found nearly the same total source age of 58 Myr, but a smaller duration for the active phase, 12 Myr instead of our 18 Myr. We note that our source age and duty cycle results are based on the resolved spectral study, which yields more precise results than the integrated one.

In the three high-resolution images between 144-675 MHz, we also detect hints of a radio core for the first time. 
This feature could represent a recent re-ignition of the SMBH or simply represent a fading low-power accretion phase. The value of the spectral index seems to point to the second possibility.

The duty cycle and active time periods that we derived for NGC 6086 are slightly shorter than the results found in other studies of radio galaxies. For example, in the B2 0924+30 remnant galaxy, \cite{shulevski2017} found an active time of $\lesssim66\%$ by performing a resolved analysis of the radio emission of a single pair of lobes. Other examples are the restarted radio galaxies in clusters 3C388 \citep{brienza2020} and CL 0838+1948 \citep{giacintucci2021} in which the authors found a lower limit of the active time of $\gtrsim 60\%$ and $\gtrsim 58\%$; in the end, \cite{biava2021} found a very rapid cycle with a short period of quiescence in the cluster central galaxy MS 0735 and they have detected an inactive time between jet outbursts of $\sim$10 Myr, less than a fifth of the active one.
The statistic of this kind of analysis is not high enough to outline differences in the duty cycle of the AGN with respect to the environment in which they are observed (clusters, groups or isolated galaxies). 

We note that the study of the duty cycle with the spectral age analysis of restarted radio galaxies is biased toward sources with rapid duty cycle, where we indeed detect multiple generations of AGN lobes. Based on statistical analyses of radio source populations, we know that these are mostly associated with massive galaxies, which often reside in rich environments ( \citealp{best2005, shabala2008, 
turner2015, sabater2019, capetti2022}). Galaxies with lower mass instead are expected to have much lower duty cycles, but for these sources, a direct estimate of the duty cycle cannot be performed because any old radio emission has faded away.

Estimates of the duty cycle of radio jets have also been performed by measuring the ages of the cavities usually seen in X-ray observations. We have no high-energy observations available to analyse the X-ray cavities of NGC 6086, but we could compare the ages we found with cavity studies carried out in other sources. High-energy studies have found a wide range of ages (25-280 Myr) when measuring the timescales related to the emissions of other radio galaxies \citep[e.g.][]{Gastaldello2009, vantyghem2014, vagshette2017, Ubertosi2021} and our measures are in good agreement with their results. Moreover, \cite{vantyghem2014} report the outburst interval of the AGN in a dozen of sources inside both groups or clusters of galaxies. They report a wide range of inactivity periods of 5-150 Myr between the outburst and our 66 and 33 Myr periods perfectly lie in the lower half of their range.

Another aspect interesting to analyse is the interaction between the radio plasma inflated by the jets and the thermal gas of the medium of groups and clusters. It is also strongly related to the duty cycle of the AGN and the cooling time of the thermal gas. This interplay could shape the radio emission in many different ways, starting with the compression or expansion of the radio lobes and ending with shaping the plasma into unique structures \citep[e.g.]{botteon2020, brienza2021, botteon2021, pandge2022}. To further complicate this scenario, AGN outbursts in clusters are thought to be more powerful and disruptive, while in group environments are expected to go through more frequent and gentle events \citep{gaspari2011}. Furthermore, interactions and mergers between systems or galaxies influence the dynamics of the plasma \citep{kolokythas2020, brienza2022}. For all these reasons, the evolution of the AGN remnant plasma in a galaxy group or cluster can be very complex. 

In the case of NGC 6086, a detailed investigation of how the outer lobes are evolving in the surrounding medium would require deep X-ray observations. However, from the analysis of their morphology and spectra, we can already make some speculations. 

In the first place, we note that the two lobes are very asymmetrical, at least in projection. The reason for the asymmetry could be the non-homogeneous IGrM in which the plasma has evolved and the consequent expansion or compression of the lobes. Another possibility is that the southern lobe is bent in the south-eastern direction in a sort of wide angle tail (WAT) scenario. \\We also note that the lobes show significant differences in the spectral indices. In particular, the spectral steepening of the eastern lobe is greater than in the western one. We hypothesised two scenarios to explain the large spectral index differences, assuming that both outer lobes formed during the same AGN outburst.\\
The first hypothesis is that the plasma in the lobes has evolved differently, resulting in different spectral shapes. For example, when strong adiabatic expansions occur, the spectrum shifts towards lower frequencies and lower intensity as it appears steeper for a given frequency range. The opposite happens in the case where the plasma is compressed. In this regard, the fact that the eastern outer lobe is steeper than the western one might be consistent with the hypothesis that it has suffered more adiabatic expansion.

Alternatively, the different spectral shapes could be justified by a different value of the magnetic field because its intensity is related to the ageing of the plasma. In our analysis, we assumed a uniform equipartition magnetic field, but if the real value is different in one of the lobes, it will affect spectral ageing and subsequent steepening. To check the consistency of this hypothesis, we considered the outer lobes to be the same age and assumed the equipartition magnetic field for the western one. Then, we inferred how the intensity of the magnetic field should be in the eastern lobe to match the age of the western one. The result is that the eastern lobe must have a magnetic field higher by a factor of 1.6 to be the same age as the western one.

Finally, we computed a first-order estimate of the non-thermal pressure inside the inner lobes. The non-thermal pressure of the plasma inside the lobes can be calculated as $\rm{P_{non-th}=u_{min}/3}$, with 
\begin{equation}
\rm{u_{min}=\frac{(3+\alpha_{inj})}{(1+\alpha_{inj})}\frac{B_{eq}^2}{8\pi}}
\end{equation}

\citep{murgia2012}. We found an equipartition value of $\rm{B_{eq}=4.4\ \mu G}$ and an injection index of $\rm{\alpha_{inj}=0.6}$ that gives a $\rm{P_{non-th}=0.6\times10^{-12}\ dyne\ cm^2}$. The non-thermal pressure found is a good estimate for the inner lobe, while for the outer lobes, it could be considered as an upper limit because, inside the outer lobes, there should be a smaller magnetic field which we do not calculate in this work.

When compared with estimates of the IGrM external pressure this can provide information on whether the radio lobes are in pressure equilibrium with the environment. To investigate the external medium and verify whether the plasma balances the thermal pressure, X-ray observations capable of providing the temperature and density profiles of the medium are needed.

\subsection{Filaments}
The high-resolution set of images revealed, for the first time in NGC 6086, three filaments of non-thermal radio emission. 
In recent years the number of filaments observed surrounding radio galaxies is rapidly increased thanks to the advent of the new generation of interferometers with high sensitivity and high dynamic range observations. The filaments are mainly found inside galaxy group or cluster environments \citep[e.g.][]{shimwell2017, ramatsoku2020, condon2021, brienza2021, rudnick2022, giacintucci2022, brienza2022}.\\

\cite{yusef2022} underline the similarities between the latest detection of radio galaxy filaments and the similar structures found in the Galactic Centre environment (i.e. morphology and spectral index). Despite the differences in size and magnetic field intensity, their similar properties could suggest that both types of filament arise through the stretching of magnetic field lines by turbulence in a weakly magnetised medium.

Overall, the origin and physics of these filaments are still under debate and it is therefore important to provide more observational constraints to inform theoretical models.

The filaments in NGC 6086 were observed in the three low-frequency images and we were able to characterise their spectrum between 144-675 MHz (see Fig. \ref{fig:filaments_fit}). When observed in the maps at high resolution, they appear as isolated structures external to the main body of the radio galaxy. However, with lower resolution we can see that, at least F1 and F2, are embedded in more diffuse emission. Interestingly, they are all oriented parallel to the main inner lobes and F1 may seem to emerge from the bent emission of the western outer lobe. This connection is seen also in \cite{giacintucci2022}, but there is still no answer as to its origin. The outer lobes are bent in the southerly direction in a WAT behaviour, the same interaction with the IGrM that generates this bending could be the cause for the stretching of the filamentary structure, as hypothesised by \cite{giacintucci2022} in the galaxy A3562.

Another anomalous filamentary stretching of the plasma is located on the eastern side of the inner lobes. In the high-resolution images, we can see a narrow thread that connects the eastern inner lobes to the eastern outer lobe emission. A similar leakage of plasma was also found by \cite{brienza2022} in the eastern lobe of NGC 507. This emission has no explanation at the moment, one possibility is that the plasma is channelled by magnetic field lines that point toward unexpected directions.

We found that the integrated spectrum of F2 is different with respect to the other two. While F1 and F3 have a power-law spectral index with comparable value within uncertainties (1.14$^{+0.13}_{-0.15}$ and 1.19$^{+0.11}_{-0.07}$, respectively), F2 shows a different behaviour with a spectral index steepening after 400 MHz ($\rm{\alpha^{400\ MHz}_{144\ MHz}=0.65\pm0.22}$ before the break and $\rm{\alpha^{675\ MHz}_{400\ MHz}=1.62\pm0.41}$ after). The reason for this spectral difference remains unclear.

Another interesting aspect is that we do not find any spectral gradient along the length of the filaments. A gradient could be expected if particles were streaming from one point to another along the filament and ageing during this path similar to what was observed in radio galaxy tails.

One plausible explanation for the formation of these filamentary structures is compression due to the interaction between the non-thermal plasma of the outer lobes with the surrounding thermal gas, possibly as a consequence of the galaxy movement or the dynamics of the group. Compression leads to an enhancement of the magnetic field intensity and particle density and to an increase in the synchrotron emission. Moreover, plasma compression should lead to a shift of the spectrum to higher frequencies. As a result, for a fixed range of frequencies, we should observe spectral flattening. Instead, we found spectral index values in the filaments which are in perfect agreement with those found in the western outer lobe between 144 MHz and 675 MHz. This suggests that if the compression is happening is not very strong.

Numerical simulations of feedback from radio galaxies suggest a potentially different scenario which can explain the spatial orientation of the filaments.
Jets launched from the centre of halos in numerical simulations typically release density and pressure perturbations even perpendicular to the  main jet axis, initially with a nearly spherical geometry, which encompasses jets \citep{bruggen2007, li2014, bourne2017}, also in accordance with earlier analytical works \citep{begelman1989}. 
Depending on the conditions of the ICM surrounding these jets, the perturbations can be seen as shocks, or weaker density and pressure waves. If such perturbations encounter a pool of relativistic electrons already present there, they can compress and re-accelerate them, generating threads of emission roughly parallel to the main jet axis, but at a perpendicular location with respect to the jet locations. Recent simulations of the interplay between electrons injected by jets and the additional ICM perturbations qualitatively support this picture \citep{vazza2023}. In this scenario, our observed filaments could have been generated thanks to the shocks produced during the jet activity of the inner lobes that have re-energized old particle populations that were already evolving in those areas.

The different spectrum of filament F2 could be consistent with this scenario because it is the result of the different evolution of the particles before the last re-energization. The spectral index values of the filaments are consistent with the values found in the western outer lobe, this evidence could indicate that the particles were inflated by the first outburst event, evolved during the first inactive phase, and finally underwent re-energization caused by the second outburst. 

The lack of filamentary structures located where the jets are directed could be caused by the higher energy of the feedback in that direction, as a consequence, parallel shocks are faster and their effects are no longer observable. However, there are cases in which filaments are still visible in the directions of the jets \citep{rudnick2022}. Moreover, evidence of perpendicular feedback with respect to the radio jets have been observed studying the direction of the enhanced gas velocity dispersion related to outflowing gas \citep[e.g.][]{riffel2015, balmaverde2019, venturi2021, ruschel2021, cuoto2023}.

Another possibility is that the filaments trace the substructures of the magnetic field in the environment. Polarisation data would be needed to further explore this aspect.

\section{Conclusions}
\label{sec:concl}
NGC 6086 has always been claimed to be a remnant radio galaxy, which presents a pair of fading lobes. In this work, we present new radio observations, which reveal new extended, filamentary emission surrounding the previously known lobes. We interpret this new emission as inflated by a previous phase of activity of the AGN. We investigate the age of the substructures and constrain the duty cycle of the radio galaxy.

Here we summarise our main findings.

\begin{itemize}
    \item [$\bullet$] Thanks to the new high-sensitivity images, we improve our knowledge of the morphology of NGC 6086, confirming the emission from the inner lobe and detecting the newly discovered diffuse emission. The full extension of the new emission is up to $\sim280$ kpc in the LOFAR image. 
    
    \item [$\bullet$] in the high-resolution set of images below 1.4 GHz, we have detected hints of a radio core and measured the flux density between 144-675 MHz. The spectrum is consistent with a power-law with spectral index $\alpha\sim1.27$ We suggest that this feature could represent a recent re-ignition of the SMBH or a fading low-power accretion phase of the SMBH. The second hypothesis is more consistent with the steep spectral index found. Future high-sensitivity and high-resolution observations are needed to investigate its spectral index and try to understand its nature.


    \item[$\bullet$] we investigated for the first time, the spectral index trends of the inner lobes between 144-675 MHz, both with resolved maps and spectral profiles. We found a mild steepening in the central region, where the host galaxy is located. A similar trend is found also between 1400-5000 MHz with steeper spectral index values. These results imply the presence of older electron populations near the host galaxy than near the edge of the lobes; \\
    
    \item[$\bullet$] the low-resolution image set was used to investigate the outer lobes of the radio galaxy. We generated a resolved spectral index map and analysed the spectral profile throughout the source. The eastern outer lobe reaches higher spectral index values than the western counterpart. Furthermore, they have a different spectral index trend as they move toward the host galaxy. The most plausible scenario is that the outer lobes undergo a different dynamical evolution, in which the environment plays a fundamental role (i.e. with adiabatic expansion, compression or variation of the magnetic field intensity);\\
    
    \item[$\bullet$] we used the \texttt{BRATS} software to perform a resolved spectral ageing analysis of the electron populations within the lobes. We used a single-injection JP model with $\rm{\alpha_{inj}=0.6}$ and two magnetic field values: the equipartition magnetic field $\rm{B_{eq}=4.4\ \mu G}$ and the lower limit $\rm{B_{min}=2.0\ \mu G}$. We derived that the last active phase which inflates the inner lobes occurred between 33-51 Myr ago with $\rm{B_{eq}}$ and between 45-69 Myr ago with $\rm{B_{min}}$;\\

    \item[$\bullet$] to provide an estimate of the age of the outer lobes, we generated a set of simulated spectra using the same injection and magnetic field used in the inner lobes analysis. Our results showed that the eastern outer lobe is more aged than the western one. The first was injected between 142-202 Myr ago with the $\rm{B_{eq}}$ value and between 274-192 with the $\rm{B_{min}}$. The latter was injected between 92-122 Myr ago with $\rm{B_{eq}}$ and between 125-166 Myr ago with $\rm{B_{min}}$;\\
    
    \item[$\bullet$] with the equipartition assumption, the duty cycle is characterised by a first active phase that started 162 Myr ago and lasted 45 Myr; this was followed by a quiescent phase of 66 Myr after which a second active phase started lasted for 18 Myr. With the minimum magnetic field assumption, the cycle is longer and characterised by a first active phase that lasted 61 Myr and started 220 Myr ago; a 90 Myr quiescent phase; a second active phase that lasted 24 Myr and started 69 Myr ago. The fraction active time of the source with $\rm{B_{eq}}$ or $\rm{B_{min}}$ is $\sim$39\%\\

    \item[$\bullet$] the three high-resolution and low-frequency images showed the presence of three new filamentary structures (F1, F2, F3), previously undetected in NGC 6086. The spectrum of F1 and F3 is a power-law between 144-675 MHZ with spectral index $1.14^{+0.13}_{-0.11}$ and $1.19^{+0.11}_{-0.07}$ respectively. The spectrum of F2 instead shows a break above 400 MHz, with alpha $0.65\pm0.22$ and $1.62\pm0.41$. 
    We speculate two possible scenarios: (i) they have been generated due to the channelling of the plasma in the magnetic field substructures or (ii) the compression of the plasma. The compression may be originated due to the gas motion inside the galaxy group or due to the shocks of the AGN formed during the second outburst. 

\end{itemize}
Future investigation with highly sensitive observations at low frequencies in the radio band will improve our knowledge of NGC 6086, for example, we could use LOFAR Low Band Antennas to investigate whether the outer lobes are more extended than currently detected, MeerKAT Radio Telescope data to detect the diffuse emission at frequencies higher than 675 MHz and collect polarization data of the filaments, to understand the role of the magnetic field in their formation. Lastly, X-ray studies could be used to look for cavities inflated by the radio jets, which could give us reliable information about the energy of the system and the jet power needed to inflate those cavities, and to look for discontinuities suggestive of gas motions. 

\begin{acknowledgements} 
AB acknowledges support from the ERC through the ERCStg 714245 DRANOEL. RJvW acknowledges support from the ERC Starting Grant ClusterWeb 804208. MB acknowledges financial support from the agreement ASI-INAF n. 2017-14-H.O, from the PRIN MIUR 2017PH3WAT “Blackout” and from the ERC-Stg `MAGCOW', no. 714196. KR acknowledges funding from Chandra grant GO0-21112X and ERC starting grant ‘MAGCOW’ number 714196. This research has made use of the NASA/IPAC Extragalactic Database (NED), which is operated by the Jet propulsion Laboratory, California Institute of Technology, under contract with the National Aeronautics and Space Administration. This research made use of APLpy, an open-source plotting package for Python hosted at \url{http://aplpy.github.com}. FL acknowledges financial support from the Italian Ministry of University and Research $-$ Project Proposal CIR01$\_$00010. FV acknowledges financial support from the H2020 St G Magcow (714196) and from the Cariplo `BREAHTHRU' funds Rif: 2022-2088 CUP J33C22004310003.

LOFAR, the Low Frequency Array designed and constructed by ASTRON, has facilities in several countries, which are owned by various parties (each with their own funding sources), and are collectively operated by the  International LOFAR Telescope (ILT) foundation under a joint scientific policy. The ILT resources have benefited from the following recent major funding sources: CNRS-INSU, Observatoire de Paris and Universit\'e d'Orl\'eans, France; BMBF, MIWF-NRW, MPG, Germany; Science Foundation Ireland (SFI), Department of Business, Enterprise and Innovation (DBEI), Ireland; NWO, The Netherlands; the Science and Technology Facilities Council, UK; Ministry of Science and Higher Education, Poland.

Part of this work was carried out on the Dutch national e-infrastructure with the support of the SURF Cooperative through grant e-infra 160022 \& 160152. The LOFAR software and dedicated reduction packages on \url{https://github.com/apmechev/GRID_LRT} were deployed on the e-infrastructure by the LOFAR e-infragroup, consisting of J.\ B.\ R.\ Oonk (ASTRON \& Leiden Observatory), A.\ P.\ Mechev (Leiden Observatory) and T. Shimwell (ASTRON) with support from N.\ Danezi (SURFsara) and C.\ Schrijvers (SURFsara). This research has made use of the University of Hertfordshire high-performance computing facility (\url{https://uhhpc.herts.ac.uk/}) and the LOFAR-UK compute facility, located at the University of Hertfordshire and supported by STFC [ST/P000096/1]. The J\"ulich LOFAR Long Term Archive and the German
LOFAR network are both coordinated and operated by the J\"ulich Supercomputing Centre (JSC), and computing resources on the supercomputer JUWELS at JSC were provided by the Gauss Centre for supercomputing e.V. (grant CHTB00) through the John von Neumann Institute for Computing (NIC).

\end{acknowledgements}

\bibliographystyle{aa}
\bibliography{ngc6086.bib}

\begin{thebibliography}{113}
\expandafter\ifx\csname natexlab\endcsname\relax\def\natexlab#1{#1}\fi

\bibitem[{{Abell} {et~al.}(1989){Abell}, {Corwin}, \& {Olowin}}]{abell1989}
{Abell}, G.~O., {Corwin}, Harold~G., J., \& {Olowin}, R.~P. 1989, \apjs, 70, 1

\bibitem[{{Balmaverde} {et~al.}(2019){Balmaverde}, {Capetti}, {Marconi},
  {Venturi}, {Chiaberge}, {Baldi}, {Baum}, {Gilli}, {Grandi}, {Meyer}, {Miley},
  {O'Dea}, {Sparks}, {Torresi}, \& {Tremblay}}]{balmaverde2019}
{Balmaverde}, B., {Capetti}, A., {Marconi}, A., {et~al.} 2019, \aap, 632, A124

\bibitem[{{Begelman} \& {Cioffi}(1989)}]{begelman1989}
{Begelman}, M.~C. \& {Cioffi}, D.~F. 1989, \apjl, 345, L21

\bibitem[{{Best} {et~al.}(2005){Best}, {Kauffmann}, {Heckman}, {Brinchmann},
  {Charlot}, {Ivezi{\'c}}, \& {White}}]{best2005}
{Best}, P.~N., {Kauffmann}, G., {Heckman}, T.~M., {et~al.} 2005, \mnras, 362,
  25

\bibitem[{{Biava} {et~al.}(2021){Biava}, {Brienza}, {Bonafede}, {Gitti},
  {Bonnassieux}, {Harwood}, {Edge}, {Riseley}, \& {Vantyghem}}]{biava2021}
{Biava}, N., {Brienza}, M., {Bonafede}, A., {et~al.} 2021, Astronomy and
  Astrophysics, 650, A170

\bibitem[{{Botteon} {et~al.}(2020){Botteon}, {Brunetti}, {van Weeren},
  {Shimwell}, {Pizzo}, {Cassano}, {Iacobelli}, {Gastaldello}, {B{\^\i}rzan},
  {Bonafede}, {Br{\"u}ggen}, {Cuciti}, {Dallacasa}, {de Gasperin}, {Di
  Gennaro}, {Drabent}, {Hardcastle}, {Hoeft}, {Mandal}, {R{\"o}ttgering}, \&
  {Simionescu}}]{botteon2020}
{Botteon}, A., {Brunetti}, G., {van Weeren}, R.~J., {et~al.} 2020, \apj, 897,
  93

\bibitem[{{Botteon} {et~al.}(2021){Botteon}, {Giacintucci}, {Gastaldello},
  {Venturi}, {Brunetti}, {van Weeren}, {Shimwell}, {Rossetti}, {Akamatsu},
  {Br{\"u}ggen}, {Cassano}, {Cuciti}, {de Gasperin}, {Drabent}, {Hoeft},
  {Mandal}, {R{\"o}ttgering}, \& {Tasse}}]{botteon2021}
{Botteon}, A., {Giacintucci}, S., {Gastaldello}, F., {et~al.} 2021, \aap, 649,
  A37

\bibitem[{{Bourne} \& {Sijacki}(2017)}]{bourne2017}
{Bourne}, M.~A. \& {Sijacki}, D. 2017, \mnras, 472, 4707

\bibitem[{{Brienza} {et~al.}(2022){Brienza}, {Lovisari}, {Rajpurohit},
  {Bonafede}, {Gastaldello}, {Murgia}, {Vazza}, {Bonnassieux}, {Botteon},
  {Brunetti}, {Drabent}, {Hardcastle}, {Pasini}, {Riseley}, {R{\"o}ttgering},
  {Shimwell}, {Simionescu}, \& {van Weeren}}]{brienza2022}
{Brienza}, M., {Lovisari}, L., {Rajpurohit}, K., {et~al.} 2022, Astronomy and
  Astrophysics, 661, A92, provided by the SAO/NASA Astrophysics Data System

\bibitem[{{Brienza} {et~al.}(2020){Brienza}, {Morganti}, {Harwood}, {Duchet},
  {Rajpurohit}, {Shulevski}, {Hardcastle}, {Mahatma}, {Godfrey}, {Prandoni},
  {Shimwell}, \& {Intema}}]{brienza2020}
{Brienza}, M., {Morganti}, R., {Harwood}, J., {et~al.} 2020, Astronomy and
  Astrophysics, 638, A29

\bibitem[{{Brienza} {et~al.}(2021){Brienza}, {Shimwell}, {de Gasperin},
  {Bikmaev}, {Bonafede}, {Botteon}, {Br{\"u}ggen}, {Brunetti}, {Burenin},
  {Capetti}, {Churazov}, {Hardcastle}, {Khabibullin}, {Lyskova},
  {R{\"o}ttgering}, {Sunyaev}, {van Weeren}, {Gastaldello}, {Mandal}, {Purser},
  {Simionescu}, \& {Tasse}}]{brienza2021}
{Brienza}, M., {Shimwell}, T.~W., {de Gasperin}, F., {et~al.} 2021, Nature
  Astronomy, 5, 1261

\bibitem[{{Brocksopp} {et~al.}(2007){Brocksopp}, {Kaiser}, {Schoenmakers}, \&
  {de Bruyn}}]{brocksopp2007}
{Brocksopp}, C., {Kaiser}, C.~R., {Schoenmakers}, A.~P., \& {de Bruyn}, A.~G.
  2007, \mnras, 382, 1019

\bibitem[{{Br{\"u}ggen} {et~al.}(2007){Br{\"u}ggen}, {Heinz}, {Roediger},
  {Ruszkowski}, \& {Simionescu}}]{bruggen2007}
{Br{\"u}ggen}, M., {Heinz}, S., {Roediger}, E., {Ruszkowski}, M., \&
  {Simionescu}, A. 2007, \mnras, 380, L67

\bibitem[{{Burns} {et~al.}(1994){Burns}, {Rhee}, {Owen}, \&
  {Pinkney}}]{burns1994}
{Burns}, J.~O., {Rhee}, G., {Owen}, F.~N., \& {Pinkney}, J. 1994, The
  Astrophysical Journal, 423, 94

\bibitem[{{Capetti} {et~al.}(2022){Capetti}, {Brienza}, {Balmaverde}, {Best},
  {Baldi}, {Drabent}, {G{\"u}rkan}, {Rottgering}, {Tasse}, \&
  {Webster}}]{capetti2022}
{Capetti}, A., {Brienza}, M., {Balmaverde}, B., {et~al.} 2022, \aap, 660, A93

\bibitem[{{Chandra} {et~al.}(2004){Chandra}, {Ray}, \&
  {Bhatnagar}}]{chandra2004}
{Chandra}, P., {Ray}, A., \& {Bhatnagar}, S. 2004, \apj, 612, 974

\bibitem[{{Colla} {et~al.}(1970){Colla}, {Fanti}, {Ficarra}, {Formiggini},
  {Gandolfi}, {Grueff}, {Lari}, {Padrielli}, {Roffi}, {Tomasi}, \&
  {Vigotti}}]{colla1970}
{Colla}, G., {Fanti}, C., {Ficarra}, A., {et~al.} 1970, \aaps, 1, 281

\bibitem[{Condon(1992)}]{condon1992}
Condon, J.~J. 1992, Annual Review of Astronomy and Astrophysics, 30, 575

\bibitem[{{Condon} {et~al.}(2021){Condon}, {Cotton}, {White}, {Legodi},
  {Goedhart}, {McAlpine}, {Ratcliffe}, \& {Camilo}}]{condon2021}
{Condon}, J.~J., {Cotton}, W.~D., {White}, S.~V., {et~al.} 2021, \apj, 917, 18

\bibitem[{{Couto} \& {Storchi-Bergmann}(2023)}]{cuoto2023}
{Couto}, G.~S. \& {Storchi-Bergmann}, T. 2023, Galaxies, 11, 47

\bibitem[{{Croston} {et~al.}(2018){Croston}, {Ineson}, \&
  {Hardcastle}}]{croston2018}
{Croston}, J.~H., {Ineson}, J., \& {Hardcastle}, M.~J. 2018, \mnras, 476, 1614

\bibitem[{{Dabhade} {et~al.}(2020){Dabhade}, {R{\"o}ttgering}, {Bagchi},
  {Shimwell}, {Hardcastle}, {Sankhyayan}, {Morganti}, {Jamrozy}, {Shulevski},
  \& {Duncan}}]{dabhade2020}
{Dabhade}, P., {R{\"o}ttgering}, H.~J.~A., {Bagchi}, J., {et~al.} 2020, \aap,
  635, A5

\bibitem[{{de Gasperin} {et~al.}(2019){de Gasperin}, {Dijkema}, {Drabent},
  {Mevius}, {Rafferty}, {van Weeren}, {Br{\"u}ggen}, {Callingham}, {Emig},
  {Heald}, {Intema}, {Morabito}, {Offringa}, {Oonk}, {Orr{\`u}},
  {R{\"o}ttgering}, {Sabater}, {Shimwell}, {Shulevski}, \&
  {Williams}}]{degasperin2019}
{de Gasperin}, F., {Dijkema}, T.~J., {Drabent}, A., {et~al.} 2019, \aap, 622,
  A5

\bibitem[{{Eckert} {et~al.}(2021){Eckert}, {Gaspari}, {Gastaldello}, {Le Brun},
  \& {O'Sullivan}}]{eckert2021}
{Eckert}, D., {Gaspari}, M., {Gastaldello}, F., {Le Brun}, A. M.~C., \&
  {O'Sullivan}, E. 2021, Universe, 7, 142

\bibitem[{{Einasto} {et~al.}(2001){Einasto}, {Einasto}, {Tago}, {M{\"u}ller},
  \& {Andernach}}]{einasto2001}
{Einasto}, M., {Einasto}, J., {Tago}, E., {M{\"u}ller}, V., \& {Andernach}, H.
  2001, \aj, 122, 2222

\bibitem[{{Fabian} {et~al.}(2005){Fabian}, {Reynolds}, {Taylor}, \&
  {Dunn}}]{fabian2005}
{Fabian}, A.~C., {Reynolds}, C.~S., {Taylor}, G.~B., \& {Dunn}, R.~J.~H. 2005,
  \mnras, 363, 891

\bibitem[{{Gaspari} {et~al.}(2011){Gaspari}, {Brighenti}, {D'Ercole}, \&
  {Melioli}}]{gaspari2011}
{Gaspari}, M., {Brighenti}, F., {D'Ercole}, A., \& {Melioli}, C. 2011, \mnras,
  415, 1549

\bibitem[{{Gastaldello} {et~al.}(2009){Gastaldello}, {Buote}, {Temi},
  {Brighenti}, {Mathews}, \& {Ettori}}]{Gastaldello2009}
{Gastaldello}, F., {Buote}, D.~A., {Temi}, P., {et~al.} 2009, \apj, 693, 43

\bibitem[{{Giacintucci} {et~al.}(2021){Giacintucci}, {Clarke}, {Kassim},
  {Peters}, \& {Polisensky}}]{giacintucci2021}
{Giacintucci}, S., {Clarke}, T., {Kassim}, N.~E., {Peters}, W., \&
  {Polisensky}, E. 2021, Galaxies, 9, 108

\bibitem[{{Giacintucci} {et~al.}(2012){Giacintucci}, {O'Sullivan}, {Clarke},
  {Murgia}, {Vrtilek}, {Venturi}, {David}, {Raychaudhury}, \&
  {Athreya}}]{giacintucci2012}
{Giacintucci}, S., {O'Sullivan}, E., {Clarke}, T.~E., {et~al.} 2012, \apj, 755,
  172

\bibitem[{{Giacintucci} {et~al.}(2022){Giacintucci}, {Venturi}, {Markevitch},
  {Bourdin}, {Mazzotta}, {Merluzzi}, {Dallacasa}, {Bardelli}, {Sikhosana},
  {Smirnov}, \& {Bernardi}}]{giacintucci2022}
{Giacintucci}, S., {Venturi}, T., {Markevitch}, M., {et~al.} 2022, \apj, 934,
  49

\bibitem[{{Giacintucci} {et~al.}(2007){Giacintucci}, {Venturi}, {Murgia},
  {Dallacasa}, {Athreya}, {Bardelli}, {Mazzotta}, \&
  {Saikia}}]{giacintucci2007}
{Giacintucci}, S., {Venturi}, T., {Murgia}, M., {et~al.} 2007, \aap, 476, 99

\bibitem[{Gupta {et~al.}(2017)Gupta, Ajithkumar, Kale, Nayak, Sabhapathy,
  Sureshkumar, Swami, Chengalur, Ghosh, Ishwara-Chandra, Joshi, Kanekar, Lal,
  \& Roy}]{gupta2017}
Gupta, Y., Ajithkumar, B., Kale, H., {et~al.} 2017, Current Science, 113, 707

\bibitem[{{G{\"u}rkan} {et~al.}(2022){G{\"u}rkan}, {Prandoni}, {O'Brien},
  {Raja}, {Marchetti}, {Vaccari}, {Driver}, {Taylor}, {Franzen}, {Brown},
  {Shabala}, {Andernach}, {Hopkins}, {Norris}, {Leahy}, {Bilicki},
  {Farajollahi}, {Galvin}, {Heald}, {Koribalski}, {An}, \&
  {Warhurst}}]{gurkan2022}
{G{\"u}rkan}, G., {Prandoni}, I., {O'Brien}, A., {et~al.} 2022, \mnras, 512,
  6104

\bibitem[{{Hardcastle} \& {Croston}(2020)}]{hardcastle2020}
{Hardcastle}, M.~J. \& {Croston}, J.~H. 2020, \nar, 88, 101539

\bibitem[{{Harwood} {et~al.}(2015){Harwood}, {Hardcastle}, \&
  {Croston}}]{harwood2015}
{Harwood}, J.~J., {Hardcastle}, M.~J., \& {Croston}, J.~H. 2015, \mnras, 454,
  3403

\bibitem[{{Harwood} {et~al.}(2013){Harwood}, {Hardcastle}, {Croston}, \&
  {Goodger}}]{harwood2013}
{Harwood}, J.~J., {Hardcastle}, M.~J., {Croston}, J.~H., \& {Goodger}, J.~L.
  2013, \mnras, 435, 3353

\bibitem[{{Harwood} {et~al.}(2017){Harwood}, {Hardcastle}, {Morganti},
  {Croston}, {Br{\"u}ggen}, {Brunetti}, {R{\"o}ttgering}, {Shulevski}, \&
  {White}}]{harwood2017}
{Harwood}, J.~J., {Hardcastle}, M.~J., {Morganti}, R., {et~al.} 2017, \mnras,
  469, 639

\bibitem[{{Intema}(2014)}]{intema2014}
{Intema}, H.~T. 2014, in Astronomical Society of India Conference Series,
  Vol.~13, Astronomical Society of India Conference Series

\bibitem[{{Intema} {et~al.}(2017){Intema}, {Jagannathan}, {Mooley}, \&
  {Frail}}]{intema2017}
{Intema}, H.~T., {Jagannathan}, P., {Mooley}, K.~P., \& {Frail}, D.~A. 2017,
  \aap, 598, A78

\bibitem[{{Jaffe} \& {Perola}(1973)}]{jaffe1973}
{Jaffe}, W.~J. \& {Perola}, G.~C. 1973, \aap, 26, 423

\bibitem[{{Jamrozy} {et~al.}(2007){Jamrozy}, {Konar}, {Saikia}, {Stawarz},
  {Mack}, \& {Siemiginowska}}]{jamrozy2007}
{Jamrozy}, M., {Konar}, C., {Saikia}, D.~J., {et~al.} 2007, \mnras, 378, 581

\bibitem[{{Jurlin} {et~al.}(2021){Jurlin}, {Brienza}, {Morganti}, {Wadadekar},
  {Ishwara-Chandra}, {Maddox}, \& {Mahatma}}]{jurlin2021}
{Jurlin}, N., {Brienza}, M., {Morganti}, R., {et~al.} 2021, \aap, 653, A110

\bibitem[{{Jurlin} {et~al.}(2020){Jurlin}, {Morganti}, {Brienza}, {Mandal},
  {Maddox}, {Duncan}, {Shabala}, {Hardcastle}, {Prandoni}, {R{\"o}ttgering},
  {Mahatma}, {Best}, {Mingo}, {Sabater}, {Shimwell}, \& {Tasse}}]{jurlin2020}
{Jurlin}, N., {Morganti}, R., {Brienza}, M., {et~al.} 2020, arXiv e-prints,
  arXiv:2004.09118

\bibitem[{{Katz-Stone} {et~al.}(1993){Katz-Stone}, {Rudnick}, \&
  {Anderson}}]{katzstone1993}
{Katz-Stone}, D.~M., {Rudnick}, L., \& {Anderson}, M.~C. 1993, \apj, 407, 549

\bibitem[{{Kolokythas} {et~al.}(2020){Kolokythas}, {O'Sullivan}, {Giacintucci},
  {Worrall}, {Birkinshaw}, {Raychaudhury}, {Horellou}, {Intema}, \&
  {Loubser}}]{kolokythas2020}
{Kolokythas}, K., {O'Sullivan}, E., {Giacintucci}, S., {et~al.} 2020, \mnras,
  496, 1471

\bibitem[{{Kolokythas} {et~al.}(2019){Kolokythas}, {O'Sullivan}, {Intema},
  {Raychaudhury}, {Babul}, {Giacintucci}, \& {Gitti}}]{kolokythas2019}
{Kolokythas}, K., {O'Sullivan}, E., {Intema}, H., {et~al.} 2019, \mnras, 489,
  2488

\bibitem[{{Komissarov} \& {Gubanov}(1994)}]{komissarov1994}
{Komissarov}, S.~S. \& {Gubanov}, A.~G. 1994, \aap, 285, 27

\bibitem[{{Konar} \& {Hardcastle}(2013)}]{konar2013}
{Konar}, C. \& {Hardcastle}, M.~J. 2013, \mnras, 436, 1595

\bibitem[{{Konar} {et~al.}(2006){Konar}, {Saikia}, {Jamrozy}, \&
  {Machalski}}]{konar2006}
{Konar}, C., {Saikia}, D.~J., {Jamrozy}, M., \& {Machalski}, J. 2006, \mnras,
  372, 693

\bibitem[{{Kukreti} {et~al.}(2022){Kukreti}, {Morganti}, {Shimwell},
  {Morabito}, {Beswick}, {Brienza}, {Hardcastle}, {Sweijen}, {Jackson},
  {Miley}, {Moldon}, {Oosterloo}, \& {de Gasperin}}]{kukreti2022}
{Kukreti}, P., {Morganti}, R., {Shimwell}, T.~W., {et~al.} 2022, \aap, 658, A6

\bibitem[{{Lagan{\'a}} {et~al.}(2013){Lagan{\'a}}, {Martinet}, {Durret}, {Lima
  Neto}, {Maughan}, \& {Zhang}}]{lagana2013}
{Lagan{\'a}}, T.~F., {Martinet}, N., {Durret}, F., {et~al.} 2013, \aap, 555,
  A66

\bibitem[{{Laing} {et~al.}(2008){Laing}, {Bridle}, {Parma}, {Feretti},
  {Giovannini}, {Murgia}, \& {Perley}}]{laing2008}
{Laing}, R.~A., {Bridle}, A.~H., {Parma}, P., {et~al.} 2008, \mnras, 386, 657

\bibitem[{{Li} \& {Bryan}(2014)}]{li2014}
{Li}, Y. \& {Bryan}, G.~L. 2014, \apj, 789, 54

\bibitem[{{Liuzzo} {et~al.}(2010){Liuzzo}, {Giovannini}, {Giroletti}, \&
  {Taylor}}]{liuzzo2010}
{Liuzzo}, E., {Giovannini}, G., {Giroletti}, M., \& {Taylor}, G.~B. 2010, \aap,
  516, A1

\bibitem[{{Maccagni} {et~al.}(2020){Maccagni}, {Murgia}, {Serra}, {Govoni},
  {Morokuma-Matsui}, {Kleiner}, {Buchner}, {J{\'o}zsa}, {Kamphuis},
  {Makhathini}, {Moln{\'a}r}, {Prokhorov}, {Ramaila}, {Ramatsoku}, {Thorat}, \&
  {Smirnov}}]{maccagni2020}
{Maccagni}, F.~M., {Murgia}, M., {Serra}, P., {et~al.} 2020, \aap, 634, A9

\bibitem[{{Mahatma} {et~al.}(2018){Mahatma}, {Hardcastle}, {Williams},
  {Brienza}, {Br{\"u}ggen}, {Croston}, {Gurkan}, {Harwood},
  {Kunert-Bajraszewska}, {Morganti}, {R{\"o}ttgering}, {Shimwell}, \&
  {Tasse}}]{mahatma2018}
{Mahatma}, V.~H., {Hardcastle}, M.~J., {Williams}, W.~L., {et~al.} 2018,
  \mnras, 475, 4557

\bibitem[{{Mahdavi} {et~al.}(1997){Mahdavi}, {B{\"o}hringer}, {Geller}, \&
  {Ramella}}]{mahdavi1997}
{Mahdavi}, A., {B{\"o}hringer}, H., {Geller}, M.~J., \& {Ramella}, M. 1997,
  \apj, 483, 68

\bibitem[{{Marecki} {et~al.}(2006){Marecki}, {Thomasson}, {Mack}, \&
  {Kunert-Bajraszewska}}]{marecki2006}
{Marecki}, A., {Thomasson}, P., {Mack}, K.~H., \& {Kunert-Bajraszewska}, M.
  2006, \aap, 448, 479

\bibitem[{{McMullin} {et~al.}(2007){McMullin}, {Waters}, {Schiebel}, {Young},
  \& {Golap}}]{mcmullin2007}
{McMullin}, J.~P., {Waters}, B., {Schiebel}, D., {Young}, W., \& {Golap}, K.
  2007, in Astronomical Society of the Pacific Conference Series, Vol. 376,
  Astronomical Data Analysis Software and Systems XVI, ed. R.~A. {Shaw},
  F.~{Hill}, \& D.~J. {Bell}, 127

\bibitem[{{Murgia} {et~al.}(2012){Murgia}, {Markevitch}, {Govoni}, {Parma},
  {Fanti}, {de Ruiter}, \& {Mack}}]{murgia2012}
{Murgia}, M., {Markevitch}, M., {Govoni}, F., {et~al.} 2012, \aap, 548, A75

\bibitem[{{Murgia} {et~al.}(2011){Murgia}, {Parma}, {Mack}, {de Ruiter},
  {Fanti}, {Govoni}, {Tarchi}, {Giacintucci}, \& {Markevitch}}]{murgia2011}
{Murgia}, M., {Parma}, P., {Mack}, K.-H., {et~al.} 2011, \aap, 526, A148

\bibitem[{{Nandi} \& {Saikia}(2012)}]{nandi2012}
{Nandi}, S. \& {Saikia}, D.~J. 2012, Bulletin of the Astronomical Society of
  India, 40, 121

\bibitem[{{O'Dea} \& {Saikia}(2021)}]{odea2021}
{O'Dea}, C.~P. \& {Saikia}, D.~J. 2021, \aapr, 29, 3

\bibitem[{{Oei} {et~al.}(2022){Oei}, {van Weeren}, {Hardcastle}, {Botteon},
  {Shimwell}, {Dabhade}, {Gast}, {R{\"o}ttgering}, {Br{\"u}ggen}, {Tasse},
  {Williams}, \& {Shulevski}}]{oei2022}
{Oei}, M. S.~S.~L., {van Weeren}, R.~J., {Hardcastle}, M.~J., {et~al.} 2022,
  \aap, 660, A2

\bibitem[{{Offringa} {et~al.}(2014){Offringa}, {McKinley}, {Hurley-Walker},
  {Briggs}, {Wayth}, {Kaplan}, {Bell}, {Feng}, {Neben}, {Hughes}, {Rhee},
  {Murphy}, {Bhat}, {Bernardi}, {Bowman}, {Cappallo}, {Corey}, {Deshpande},
  {Emrich}, {Ewall-Wice}, {Gaensler}, {Goeke}, {Greenhill}, {Hazelton},
  {Hindson}, {Johnston-Hollitt}, {Jacobs}, {Kasper}, {Kratzenberg}, {Lenc},
  {Lonsdale}, {Lynch}, {McWhirter}, {Mitchell}, {Morales}, {Morgan},
  {Kudryavtseva}, {Oberoi}, {Ord}, {Pindor}, {Procopio}, {Prabu}, {Riding},
  {Roshi}, {Shankar}, {Srivani}, {Subrahmanyan}, {Tingay}, {Waterson},
  {Webster}, {Whitney}, {Williams}, \& {Williams}}]{offringa2014}
{Offringa}, A.~R., {McKinley}, B., {Hurley-Walker}, N., {et~al.} 2014, \mnras,
  444, 606

\bibitem[{{Orr{\`u}} {et~al.}(2010){Orr{\`u}}, {Murgia}, {Feretti}, {Govoni},
  {Giovannini}, {Lane}, {Kassim}, \& {Paladino}}]{orru2010}
{Orr{\`u}}, E., {Murgia}, M., {Feretti}, L., {et~al.} 2010, \aap, 515, A50

\bibitem[{{Owen} \& {Ledlow}(1997)}]{owen1997}
{Owen}, F.~N. \& {Ledlow}, M.~J. 1997, The Astrophysical Journals, 108, 41

\bibitem[{{Pandge} {et~al.}(2022){Pandge}, {Kale}, {Dabhade}, {Mahato}, \&
  {Raychaudhury}}]{pandge2022}
{Pandge}, M.~B., {Kale}, R., {Dabhade}, P., {Mahato}, M., \& {Raychaudhury}, S.
  2022, \mnras, 509, 1837

\bibitem[{{Parma} {et~al.}(1986){Parma}, {de Ruiter}, {Fanti}, \&
  {Fanti}}]{parma1986}
{Parma}, P., {de Ruiter}, H.~R., {Fanti}, C., \& {Fanti}, R. 1986, \aaps, 64,
  135

\bibitem[{{Parma} {et~al.}(2007){Parma}, {Murgia}, {de Ruiter}, {Fanti},
  {Mack}, \& {Govoni}}]{parma2007}
{Parma}, P., {Murgia}, M., {de Ruiter}, H.~R., {et~al.} 2007, \aap, 470, 875

\bibitem[{{Perley} \& {Butler}(2013)}]{perley2013}
{Perley}, R.~A. \& {Butler}, B.~J. 2013, \apjs, 206, 16

\bibitem[{{Ramatsoku} {et~al.}(2020){Ramatsoku}, {Murgia}, {Vacca}, {Serra},
  {Makhathini}, {Govoni}, {Smirnov}, {Andati}, {de Blok}, {J{\'o}zsa},
  {Kamphuis}, {Kleiner}, {Maccagni}, {Moln{\'a}r}, {Ramaila}, {Thorat}, \&
  {White}}]{ramatsoku2020}
{Ramatsoku}, M., {Murgia}, M., {Vacca}, V., {et~al.} 2020, \aap, 636, L1

\bibitem[{{Randall} {et~al.}(2015){Randall}, {Nulsen}, {Jones}, {Forman},
  {Bulbul}, {Clarke}, {Kraft}, {Blanton}, {David}, {Werner}, {Sun}, {Donahue},
  {Giacintucci}, \& {Simionescu}}]{randall2015}
{Randall}, S.~W., {Nulsen}, P.~E.~J., {Jones}, C., {et~al.} 2015, \apj, 805,
  112

\bibitem[{{Rao} {et~al.}(2023){Rao}, {Kharb}, {Rubinur}, {Silpa}, {Roy},
  {Sebastian}, {Singh}, {Baghel}, {Manna}, \& {Ishwara-Chandra}}]{rao2023}
{Rao}, V.~V., {Kharb}, P., {Rubinur}, K., {et~al.} 2023, arXiv e-prints,
  arXiv:2301.01610

\bibitem[{{Rengelink} {et~al.}(1997){Rengelink}, {Tang}, {de Bruyn}, {Miley},
  {Bremer}, {Roettgering}, \& {Bremer}}]{rengelink1997}
{Rengelink}, R.~B., {Tang}, Y., {de Bruyn}, A.~G., {et~al.} 1997, \aaps, 124,
  259

\bibitem[{{Riffel} {et~al.}(2015){Riffel}, {Storchi-Bergmann}, \&
  {Riffel}}]{riffel2015}
{Riffel}, R.~A., {Storchi-Bergmann}, T., \& {Riffel}, R. 2015, \mnras, 451,
  3587

\bibitem[{{Riley}(1989)}]{riley1989}
{Riley}, J.~M. 1989, \mnras, 238, 1055

\bibitem[{{Romano} {et~al.}(2014){Romano}, {Guidorzi}, {Segreto}, {Ducci}, \&
  {Vercellone}}]{romano2014}
{Romano}, P., {Guidorzi}, C., {Segreto}, A., {Ducci}, L., \& {Vercellone}, S.
  2014, Astronomy and Astrophysics, 572, A97, provided by the SAO/NASA
  Astrophysics Data System

\bibitem[{{Rudnick} {et~al.}(2022){Rudnick}, {Br{\"u}ggen}, {Brunetti},
  {Cotton}, {Forman}, {Jones}, {Nolting}, {Schellenberger}, \& {van
  Weeren}}]{rudnick2022}
{Rudnick}, L., {Br{\"u}ggen}, M., {Brunetti}, G., {et~al.} 2022, \apj, 935, 168

\bibitem[{{Ruschel-Dutra} {et~al.}(2021){Ruschel-Dutra}, {Storchi-Bergmann},
  {Schnorr-M{\"u}ller}, {Riffel}, {Dall'Agnol de Oliveira}, {Lena}, {Robinson},
  {Nagar}, \& {Elvis}}]{ruschel2021}
{Ruschel-Dutra}, D., {Storchi-Bergmann}, T., {Schnorr-M{\"u}ller}, A., {et~al.}
  2021, \mnras, 507, 74

\bibitem[{{Sabater} {et~al.}(2019){Sabater}, {Best}, {Hardcastle}, {Shimwell},
  {Tasse}, {Williams}, {Br{\"u}ggen}, {Cochrane}, {Croston}, \& {de
  Gasperin}}]{sabater2019}
{Sabater}, J., {Best}, P.~N., {Hardcastle}, M.~J., {et~al.} 2019, \aap, 622,
  A17

\bibitem[{{Saripalli} {et~al.}(1986){Saripalli}, {Gopal-Krishna}, {Reich}, \&
  {Kuehr}}]{saripalli1986}
{Saripalli}, L., {Gopal-Krishna}, {Reich}, W., \& {Kuehr}, H. 1986, \aap, 170,
  20

\bibitem[{{Scaife} \& {Heald}(2012)}]{scaife2012}
{Scaife}, A.~M.~M. \& {Heald}, G.~H. 2012, \mnras, 423, L30

\bibitem[{{Schellenberger} {et~al.}(2021){Schellenberger}, {David}, {Vrtilek},
  {O'Sullivan}, {Giacintucci}, {Forman}, {Jones}, \&
  {Venturi}}]{schellenberger2021}
{Schellenberger}, G., {David}, L.~P., {Vrtilek}, J., {et~al.} 2021, \apj, 906,
  16

\bibitem[{{Schellenberger} {et~al.}(2023){Schellenberger}, {O'Sullivan},
  {Giacintucci}, {Vrtilek}, {David}, {Combes}, {B{\^\i}rzan}, {Pan}, \&
  {Lin}}]{Schellenberger2023}
{Schellenberger}, G., {O'Sullivan}, E., {Giacintucci}, S., {et~al.} 2023, arXiv
  e-prints, arXiv:2303.08833

\bibitem[{{Schoenmakers} {et~al.}(2000{\natexlab{a}}){Schoenmakers}, {de
  Bruyn}, {R{\"o}ttgering}, \& {van der Laan}}]{schoenmakers2000b}
{Schoenmakers}, A.~P., {de Bruyn}, A.~G., {R{\"o}ttgering}, H.~J.~A., \& {van
  der Laan}, H. 2000{\natexlab{a}}, \mnras, 315, 395

\bibitem[{{Schoenmakers} {et~al.}(2000{\natexlab{b}}){Schoenmakers}, {de
  Bruyn}, {R{\"o}ttgering}, {van der Laan}, \& {Kaiser}}]{schoenmakers2000}
{Schoenmakers}, A.~P., {de Bruyn}, A.~G., {R{\"o}ttgering}, H.~J.~A., {van der
  Laan}, H., \& {Kaiser}, C.~R. 2000{\natexlab{b}}, \mnras, 315, 371

\bibitem[{{Shabala} {et~al.}(2008){Shabala}, {Ash}, {Alexander}, \&
  {Riley}}]{shabala2008}
{Shabala}, S.~S., {Ash}, S., {Alexander}, P., \& {Riley}, J.~M. 2008, \mnras,
  388, 625

\bibitem[{{Shimwell} {et~al.}(2022){Shimwell}, {Hardcastle}, {Tasse}, {Best},
  {R{\"o}ttgering}, {Williams}, {Botteon}, {Drabent}, {Mechev}, {Shulevski},
  {van Weeren}, {Bester}, {Br{\"u}ggen}, {Brunetti}, {Callingham}, {Chy{\.z}y},
  {Conway}, {Dijkema}, {Duncan}, {de Gasperin}, {Hale}, {Haverkorn}, {Hugo},
  {Jackson}, {Mevius}, {Miley}, {Morabito}, {Morganti}, {Offringa}, {Oonk},
  {Rafferty}, {Sabater}, {Smith}, {Schwarz}, {Smirnov}, {O'Sullivan},
  {Vedantham}, {White}, {Albert}, {Alegre}, {Asabere}, {Bacon}, {Bonafede},
  {Bonnassieux}, {Brienza}, {Bilicki}, {Bonato}, {Calistro Rivera}, {Cassano},
  {Cochrane}, {Croston}, {Cuciti}, {Dallacasa}, {Danezi}, {Dettmar}, {Di
  Gennaro}, {Edler}, {En{\ss}lin}, {Emig}, {Franzen}, {Garc{\'\i}a-Vergara},
  {Grange}, {G{\"u}rkan}, {Hajduk}, {Heald}, {Heesen}, {Hoang}, {Hoeft},
  {Horellou}, {Iacobelli}, {Jamrozy}, {Jeli{\'c}}, {Kondapally}, {Kukreti},
  {Kunert-Bajraszewska}, {Magliocchetti}, {Mahatma}, {Ma{\l}ek}, {Mandal},
  {Massaro}, {Meyer-Zhao}, {Mingo}, {Mostert}, {Nair}, {Nakoneczny},
  {Nikiel-Wroczy{\'n}ski}, {Orr{\'u}}, {Pajdosz-{\'S}mierciak}, {Pasini},
  {Prandoni}, {van Piggelen}, {Rajpurohit}, {Retana-Montenegro}, {Riseley},
  {Rowlinson}, {Saxena}, {Schrijvers}, {Sweijen}, {Siewert}, {Timmerman},
  {Vaccari}, {Vink}, {West}, {Wo{\l}owska}, {Zhang}, \& {Zheng}}]{shimwell2022}
{Shimwell}, T.~W., {Hardcastle}, M.~J., {Tasse}, C., {et~al.} 2022, \aap, 659,
  A1

\bibitem[{{Shimwell} {et~al.}(2017){Shimwell}, {R{\"o}ttgering}, {Best},
  {Williams}, {Dijkema}, {de Gasperin}, {Hardcastle}, {Heald}, {Hoang},
  {Horneffer}, {Intema}, {Mahony}, {Mandal}, {Mechev}, {Morabito}, {Oonk},
  {Rafferty}, {Retana-Montenegro}, {Sabater}, {Tasse}, {van Weeren},
  {Br{\"u}ggen}, {Brunetti}, {Chy{\.z}y}, {Conway}, {Haverkorn}, {Jackson},
  {Jarvis}, {McKean}, {Miley}, {Morganti}, {White}, {Wise}, {van Bemmel},
  {Beck}, {Brienza}, {Bonafede}, {Calistro Rivera}, {Cassano}, {Clarke},
  {Cseh}, {Deller}, {Drabent}, {van Driel}, {Engels}, {Falcke}, {Ferrari},
  {Fr{\"o}hlich}, {Garrett}, {Harwood}, {Heesen}, {Hoeft}, {Horellou},
  {Israel}, {Kapi{\'n}ska}, {Kunert-Bajraszewska}, {McKay}, {Mohan},
  {Orr{\'u}}, {Pizzo}, {Prandoni}, {Schwarz}, {Shulevski}, {Sipior}, {Smith},
  {Sridhar}, {Steinmetz}, {Stroe}, {Varenius}, {van der Werf}, {Zensus}, \&
  {Zwart}}]{shimwell2017}
{Shimwell}, T.~W., {R{\"o}ttgering}, H.~J.~A., {Best}, P.~N., {et~al.} 2017,
  \aap, 598, A104

\bibitem[{{Shimwell} {et~al.}(2019){Shimwell}, {Tasse}, {Hardcastle}, {Mechev},
  {Williams}, {Best}, {R{\"o}ttgering}, {Callingham}, {Dijkema}, {de Gasperin},
  {Hoang}, {Hugo}, {Mirmont}, {Oonk}, {Prandoni}, {Rafferty}, {Sabater},
  {Smirnov}, {van Weeren}, {White}, {Atemkeng}, {Bester}, {Bonnassieux},
  {Br{\"u}ggen}, {Brunetti}, {Chy{\.z}y}, {Cochrane}, {Conway}, {Croston},
  {Danezi}, {Duncan}, {Haverkorn}, {Heald}, {Iacobelli}, {Intema}, {Jackson},
  {Jamrozy}, {Jarvis}, {Lakhoo}, {Mevius}, {Miley}, {Morabito}, {Morganti},
  {Nisbet}, {Orr{\'u}}, {Perkins}, {Pizzo}, {Schrijvers}, {Smith}, {Vermeulen},
  {Wise}, {Alegre}, {Bacon}, {van Bemmel}, {Beswick}, {Bonafede}, {Botteon},
  {Bourke}, {Brienza}, {Calistro Rivera}, {Cassano}, {Clarke}, {Conselice},
  {Dettmar}, {Drabent}, {Dumba}, {Emig}, {En{\ss}lin}, {Ferrari}, {Garrett},
  {G{\'e}nova-Santos}, {Goyal}, {G{\"u}rkan}, {Hale}, {Harwood}, {Heesen},
  {Hoeft}, {Horellou}, {Jackson}, {Kokotanekov}, {Kondapally},
  {Kunert-Bajraszewska}, {Mahatma}, {Mahony}, {Mandal}, {McKean}, {Merloni},
  {Mingo}, {Miskolczi}, {Mooney}, {Nikiel-Wroczy{\'n}ski}, {O'Sullivan},
  {Quinn}, {Reich}, {Roskowi{\'n}ski}, {Rowlinson}, {Savini}, {Saxena},
  {Schwarz}, {Shulevski}, {Sridhar}, {Stacey}, {Urquhart}, {van der Wiel},
  {Varenius}, {Webster}, \& {Wilber}}]{shimwell2019}
{Shimwell}, T.~W., {Tasse}, C., {Hardcastle}, M.~J., {et~al.} 2019, \aap, 622,
  A1

\bibitem[{{Shulevski} {et~al.}(2017){Shulevski}, {Morganti}, {Harwood},
  {Barthel}, {Jamrozy}, {Brienza}, {Brunetti}, {R{\"o}ttgering}, {Murgia},
  {White}, {Croston}, \& {Br{\"u}ggen}}]{shulevski2017}
{Shulevski}, A., {Morganti}, R., {Harwood}, J.~J., {et~al.} 2017, \aap, 600,
  A65

\bibitem[{{Singh} {et~al.}(2016){Singh}, {Ishwara-Chandra}, {Kharb},
  {Srivastava}, \& {Janardhan}}]{singh2016}
{Singh}, V., {Ishwara-Chandra}, C.~H., {Kharb}, P., {Srivastava}, S., \&
  {Janardhan}, P. 2016, \apj, 826, 132

\bibitem[{{Smirnov} \& {Tasse}(2015)}]{smirnov2015}
{Smirnov}, O.~M. \& {Tasse}, C. 2015, \mnras, 449, 2668

\bibitem[{{Tasse}(2014)}]{tasse2014}
{Tasse}, C. 2014, \aap, 566, A127

\bibitem[{{Tasse} {et~al.}(2018){Tasse}, {Hugo}, {Mirmont}, {Smirnov},
  {Atemkeng}, {Bester}, {Hardcastle}, {Lakhoo}, {Perkins}, \&
  {Shimwell}}]{tasse2018}
{Tasse}, C., {Hugo}, B., {Mirmont}, M., {et~al.} 2018, \aap, 611, A87

\bibitem[{{Tasse} {et~al.}(2021){Tasse}, {Shimwell}, {Hardcastle},
  {O'Sullivan}, {van Weeren}, {Best}, {Bester}, {Hugo}, {Smirnov}, {Sabater},
  {Calistro-Rivera}, {de Gasperin}, {Morabito}, {R{\"o}ttgering}, {Williams},
  {Bonato}, {Bondi}, {Botteon}, {Br{\"u}ggen}, {Brunetti}, {Chy{\.z}y},
  {Garrett}, {G{\"u}rkan}, {Jarvis}, {Kondapally}, {Mandal}, {Prandoni},
  {Repetti}, {Retana-Montenegro}, {Schwarz}, {Shulevski}, \&
  {Wiaux}}]{tasse2021}
{Tasse}, C., {Shimwell}, T., {Hardcastle}, M.~J., {et~al.} 2021, \aap, 648, A1

\bibitem[{{Turner} \& {Shabala}(2015)}]{turner2015}
{Turner}, R.~J. \& {Shabala}, S.~S. 2015, \apj, 806, 59

\bibitem[{{Ubertosi} {et~al.}(2021){Ubertosi}, {Gitti}, {Brighenti},
  {Brunetti}, {McDonald}, {Nulsen}, {McNamara}, {Randall}, {Forman}, {Donahue},
  {Ignesti}, {Gaspari}, {Ettori}, {Feretti}, {Blanton}, {Jones}, \&
  {Calzadilla}}]{Ubertosi2021}
{Ubertosi}, F., {Gitti}, M., {Brighenti}, F., {et~al.} 2021, \apjl, 923, L25

\bibitem[{{Vagshette} {et~al.}(2017){Vagshette}, {Naik}, {Patil}, \&
  {Sonkamble}}]{vagshette2017}
{Vagshette}, N.~D., {Naik}, S., {Patil}, M.~K., \& {Sonkamble}, S.~S. 2017,
  \mnras, 466, 2054

\bibitem[{{van Haarlem} {et~al.}(2013){van Haarlem}, {Wise}, {Gunst}, {Heald},
  {McKean}, {Hessels}, {de Bruyn}, {Nijboer}, {Swinbank}, {Fallows},
  {Brentjens}, {Nelles}, {Beck}, {Falcke}, {Fender}, {H{\"o}randel},
  {Koopmans}, {Mann}, {Miley}, {R{\"o}ttgering}, {Stappers}, {Wijers},
  {Zaroubi}, {van den Akker}, {Alexov}, {Anderson}, {Anderson}, {van Ardenne},
  {Arts}, {Asgekar}, {Avruch}, {Batejat}, {B{\"a}hren}, {Bell}, {Bell}, {van
  Bemmel}, {Bennema}, {Bentum}, {Bernardi}, {Best}, {B{\^i}rzan}, {Bonafede},
  {Boonstra}, {Braun}, {Bregman}, {Breitling}, {van de Brink}, {Broderick},
  {Broekema}, {Brouw}, {Br{\"u}ggen}, {Butcher}, {van Cappellen}, {Ciardi},
  {Coenen}, {Conway}, {Coolen}, {Corstanje}, {Damstra}, {Davies}, {Deller},
  {Dettmar}, {van Diepen}, {Dijkstra}, {Donker}, {Doorduin}, {Dromer}, {Drost},
  {van Duin}, {Eisl{\"o}ffel}, {van Enst}, {Ferrari}, {Frieswijk}, {Gankema},
  {Garrett}, {de Gasperin}, {Gerbers}, {de Geus}, {Grie{\ss}meier}, {Grit},
  {Gruppen}, {Hamaker}, {Hassall}, {Hoeft}, {Holties}, {Horneffer}, {van der
  Horst}, {van Houwelingen}, {Huijgen}, {Iacobelli}, {Intema}, {Jackson},
  {Jelic}, {de Jong}, {Juette}, {Kant}, {Karastergiou}, {Koers}, {Kollen},
  {Kondratiev}, {Kooistra}, {Koopman}, {Koster}, {Kuniyoshi}, {Kramer},
  {Kuper}, {Lambropoulos}, {Law}, {van Leeuwen}, {Lemaitre}, {Loose}, {Maat},
  {Macario}, {Markoff}, {Masters}, {McFadden}, {McKay-Bukowski}, {Meijering},
  {Meulman}, {Mevius}, {Middelberg}, {Millenaar}, {Miller-Jones}, {Mohan},
  {Mol}, {Morawietz}, {Morganti}, {Mulcahy}, {Mulder}, {Munk}, {Nieuwenhuis},
  {van Nieuwpoort}, {Noordam}, {Norden}, {Noutsos}, {Offringa}, {Olofsson},
  {Omar}, {Orr{\'u}}, {Overeem}, {Paas}, {Pandey-Pommier}, {Pandey}, {Pizzo},
  {Polatidis}, {Rafferty}, {Rawlings}, {Reich}, {de Reijer}, {Reitsma},
  {Renting}, {Riemers}, {Rol}, {Romein}, {Roosjen}, {Ruiter}, {Scaife}, {van
  der Schaaf}, {Scheers}, {Schellart}, {Schoenmakers}, {Schoonderbeek},
  {Serylak}, {Shulevski}, {Sluman}, {Smirnov}, {Sobey}, {Spreeuw}, {Steinmetz},
  {Sterks}, {Stiepel}, {Stuurwold}, {Tagger}, {Tang}, {Tasse}, {Thomas},
  {Thoudam}, {Toribio}, {van der Tol}, {Usov}, {van Veelen}, {van der Veen},
  {ter Veen}, {Verbiest}, {Vermeulen}, {Vermaas}, {Vocks}, {Vogt}, {de Vos},
  {van der Wal}, {van Weeren}, {Weggemans}, {Weltevrede}, {White}, {Wijnholds},
  {Wilhelmsson}, {Wucknitz}, {Yatawatta}, {Zarka}, {Zensus}, \& {van
  Zwieten}}]{vanhaarlem2013}
{van Haarlem}, M.~P., {Wise}, M.~W., {Gunst}, A.~W., {et~al.} 2013, \aap, 556,
  A2

\bibitem[{{van Weeren} {et~al.}(2021){van Weeren}, {Shimwell}, {Botteon},
  {Brunetti}, {Br{\"u}ggen}, {Boxelaar}, {Cassano}, {Di Gennaro},
  {Andrade-Santos}, {Bonnassieux}, {Bonafede}, {Cuciti}, {Dallacasa}, {de
  Gasperin}, {Gastaldello}, {Hardcastle}, {Hoeft}, {Kraft}, {Mandal},
  {Rossetti}, {R{\"o}ttgering}, {Tasse}, \& {Wilber}}]{vanweeren2021}
{van Weeren}, R.~J., {Shimwell}, T.~W., {Botteon}, A., {et~al.} 2021, \aap,
  651, A115

\bibitem[{{van Weeren} {et~al.}(2016){van Weeren}, {Williams}, {Hardcastle},
  {Shimwell}, {Rafferty}, {Sabater}, {Heald}, {Sridhar}, {Dijkema}, {Brunetti},
  {Br{\"u}ggen}, {Andrade-Santos}, {Ogrean}, {R{\"o}ttgering}, {Dawson},
  {Forman}, {de Gasperin}, {Jones}, {Miley}, {Rudnick}, {Sarazin}, {Bonafede},
  {Best}, {B{\^i}rzan}, {Cassano}, {Chy{\.z}y}, {Croston}, {Ensslin},
  {Ferrari}, {Hoeft}, {Horellou}, {Jarvis}, {Kraft}, {Mevius}, {Intema},
  {Murray}, {Orr{\'u}}, {Pizzo}, {Simionescu}, {Stroe}, {van der Tol}, \&
  {White}}]{vanweeren2016}
{van Weeren}, R.~J., {Williams}, W.~L., {Hardcastle}, M.~J., {et~al.} 2016,
  \apjs, 223, 2

\bibitem[{{Vantyghem} {et~al.}(2014){Vantyghem}, {McNamara}, {Russell}, {Main},
  {Nulsen}, {Wise}, {Hoekstra}, \& {Gitti}}]{vantyghem2014}
{Vantyghem}, A.~N., {McNamara}, B.~R., {Russell}, H.~R., {et~al.} 2014, \mnras,
  442, 3192

\bibitem[{{Vazza} {et~al.}(2023){Vazza}, {Wittor}, {Di Federico},
  {Br{\"u}ggen}, {Brienza}, {Brunetti}, {Brighenti}, \& {Pasini}}]{vazza2023}
{Vazza}, F., {Wittor}, D., {Di Federico}, L., {et~al.} 2023, \aap, 669, A50

\bibitem[{{Venturi} {et~al.}(2021){Venturi}, {Cresci}, {Marconi}, {Mingozzi},
  {Nardini}, {Carniani}, {Mannucci}, {Marasco}, {Maiolino}, {Perna},
  {Treister}, {Bland-Hawthorn}, \& {Gallimore}}]{venturi2021}
{Venturi}, G., {Cresci}, G., {Marconi}, A., {et~al.} 2021, \aap, 648, A17

\bibitem[{{Venturi} {et~al.}(2004){Venturi}, {Dallacasa}, \&
  {Stefanachi}}]{venturi2004}
{Venturi}, T., {Dallacasa}, D., \& {Stefanachi}, F. 2004, \aap, 422, 515

\bibitem[{{Webster} {et~al.}(2021){Webster}, {Croston}, {Harwood}, {Baldi},
  {Hardcastle}, {Mingo}, \& {R{\"o}ttgering}}]{webster2021}
{Webster}, B., {Croston}, J.~H., {Harwood}, J.~J., {et~al.} 2021, \mnras, 508,
  5972

\bibitem[{{Williams} {et~al.}(2016){Williams}, {van Weeren}, {R{\"o}ttgering},
  {Best}, {Dijkema}, {de Gasperin}, {Hardcastle}, {Heald}, {Prandoni},
  {Sabater}, {Shimwell}, {Tasse}, {van Bemmel}, {Br{\"u}ggen}, {Brunetti},
  {Conway}, {En{\ss}lin}, {Engels}, {Falcke}, {Ferrari}, {Haverkorn},
  {Jackson}, {Jarvis}, {Kapi{\'n}ska}, {Mahony}, {Miley}, {Morabito},
  {Morganti}, {Orr{\'u}}, {Retana-Montenegro}, {Sridhar}, {Toribio}, {White},
  {Wise}, \& {Zwart}}]{williams2016}
{Williams}, W.~L., {van Weeren}, R.~J., {R{\"o}ttgering}, H.~J.~A., {et~al.}
  2016, \mnras, 460, 2385

\bibitem[{{Willis} {et~al.}(1974){Willis}, {Strom}, \& {Wilson}}]{willis1974}
{Willis}, A.~G., {Strom}, R.~G., \& {Wilson}, A.~S. 1974, \nat, 250, 625

\bibitem[{{Wise} {et~al.}(2007){Wise}, {McNamara}, {Nulsen}, {Houck}, \&
  {David}}]{wise2007}
{Wise}, M.~W., {McNamara}, B.~R., {Nulsen}, P.~E.~J., {Houck}, J.~C., \&
  {David}, L.~P. 2007, \apj, 659, 1153

\bibitem[{{Yusef-Zadeh} {et~al.}(2022){Yusef-Zadeh}, {Arendt}, \&
  {Wardle}}]{yusef2022}
{Yusef-Zadeh}, F., {Arendt}, R.~G., \& {Wardle}, M. 2022, \apjl, 939, L21

\end{thebibliography}

\appendix
\onecolumn
\section{Additional images}
\label{appendix}

\begin{figure*}[htp!]
    \centering
     \includegraphics[width=0.45\textwidth]{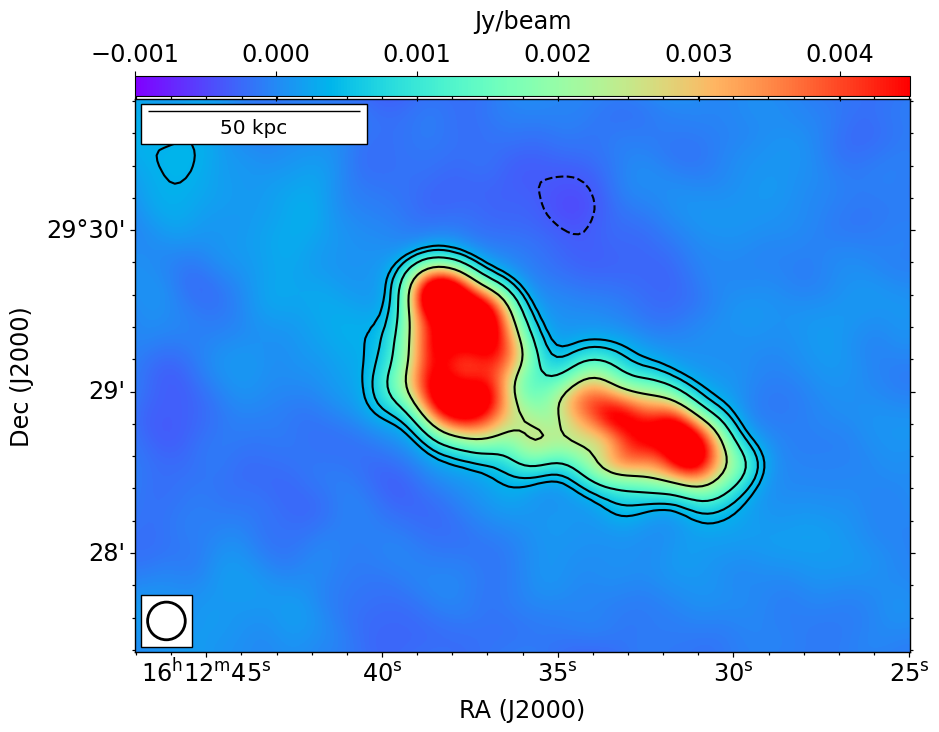}
     \includegraphics[width=0.45\textwidth]{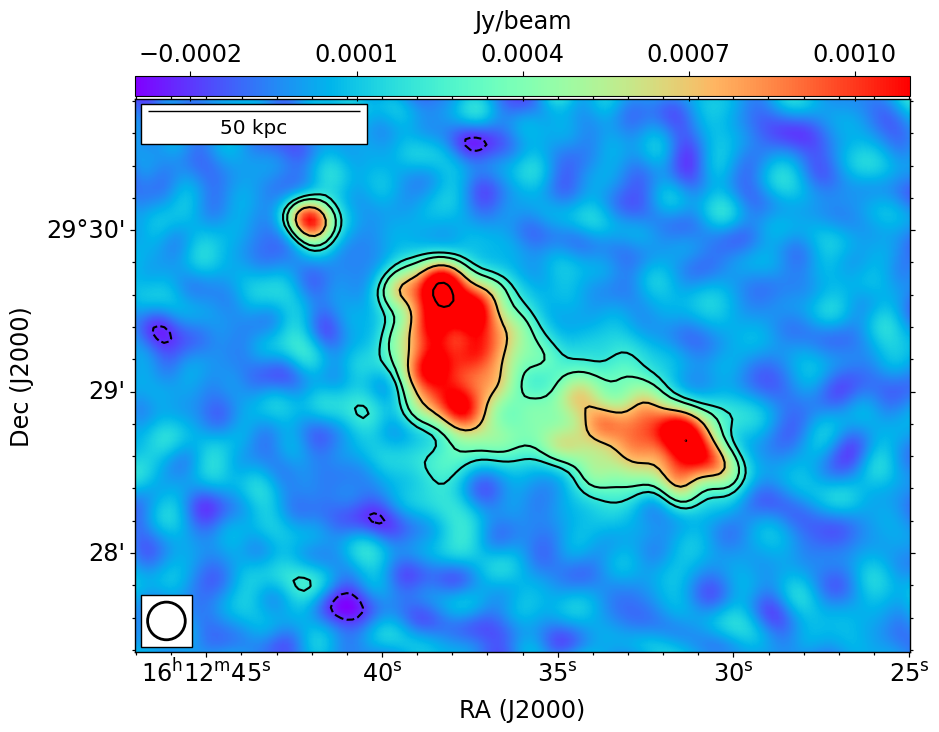}
     \caption{Radio images of the source NGC 6086 at 1400 MHz (left) and 4700 MHz (right) at 14 arcsec of resolution. Contours are drawn at (-3, 3, 5, 10, 20)$\times\sigma$. The beam is shown in the bottom-left corner and a reference physical scale is shown in the top-left corner.}
     \label{Fig:ngc6086_vla}
\end{figure*}

\begin{figure*}[htp!]
    \centering
    \includegraphics[width=0.6\textwidth]{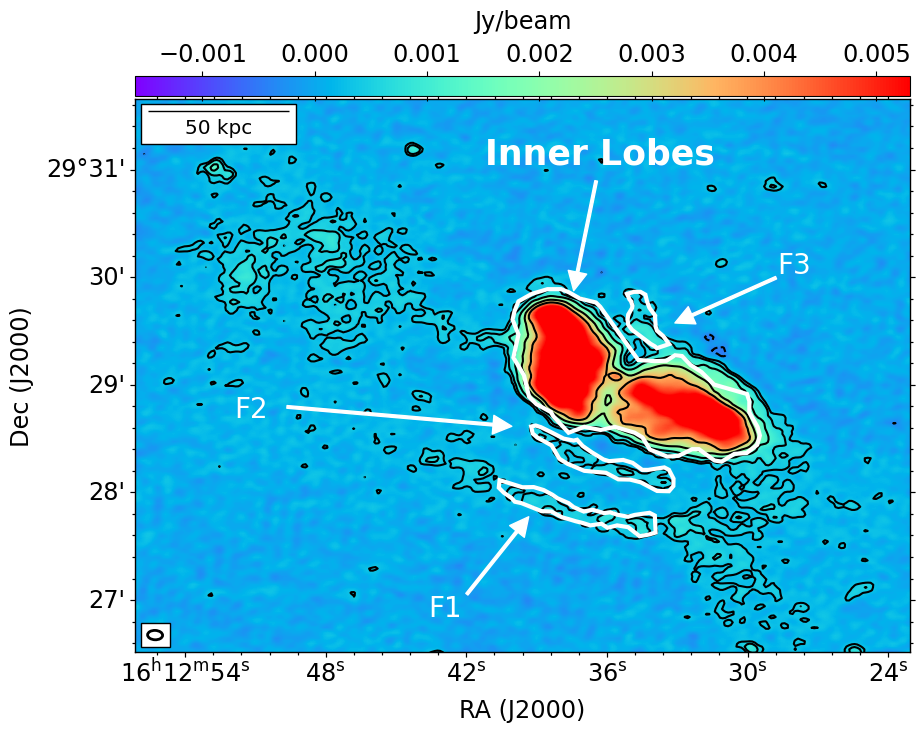}
    \caption{We highlight the radio substructures of the three filaments and the inner lobes in the radio image of the source NGC 6086 at 144 MHz at 8.4 arcsec $\times$ 5.2 arcsec of resolution. Contours are drawn at (-3, 3, 5, 10, 20, 40)$\times\sigma$. The beam is shown in the bottom-left corner and a reference physical scale is shown in the top-left corner.}
     \label{fig:substructures}
\end{figure*}

\begin{figure*}[!htp]
    \centering
        \includegraphics[width=0.41\textwidth]{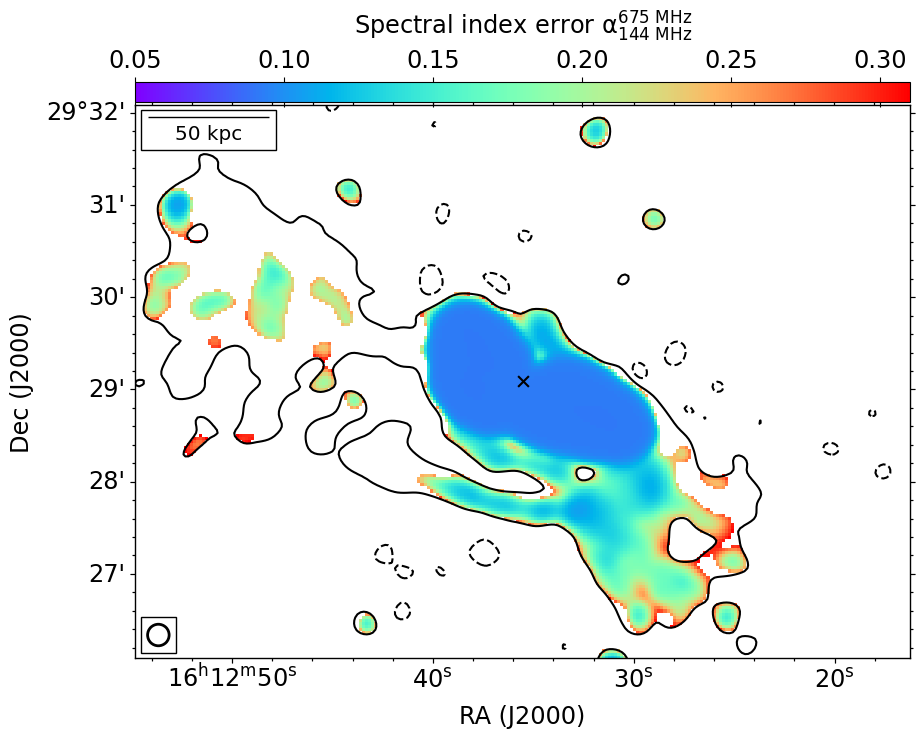}
        \includegraphics[width=0.41\textwidth]{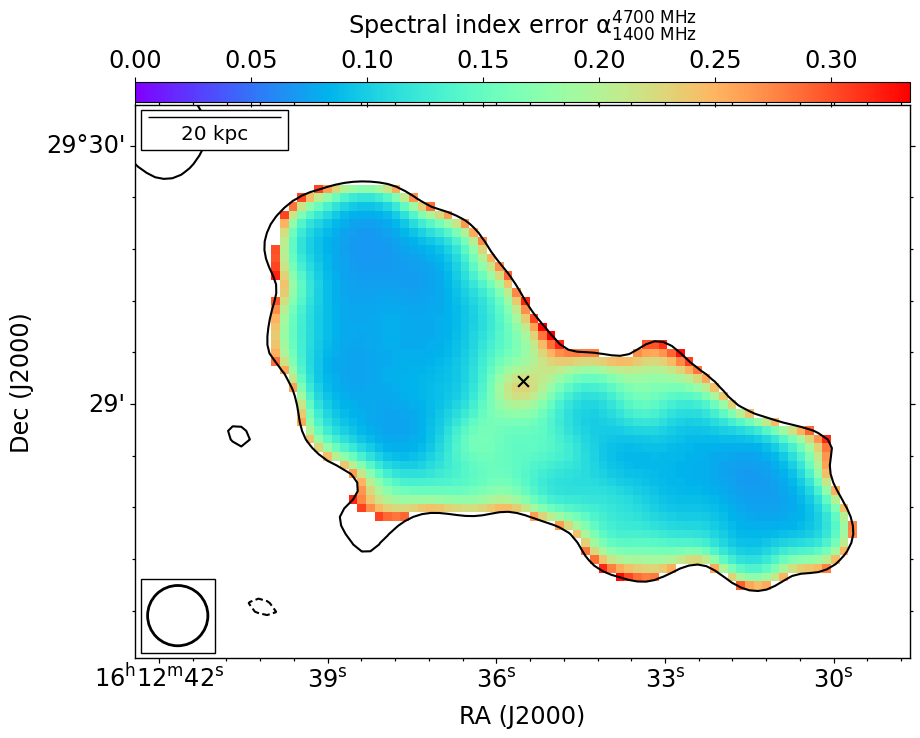}
     \caption{Left: NGC 6086 error value associated with the spectral index map at 14 arcsec between 144-675 MHz. Contours show the LOFAR emission and are drawn at 3$\sigma$. Right: NGC 6086 error value associated with the spectral index map at 14 arcsec between 1400-4700 MHz. Contours show the VLA 4.7 GHz emission and are drawn at 3$\sigma$. The beam is shown in the bottom-left corner and a reference physical scale is shown in the top-left corner.}
     \label{fig:6086_14_error_maps}
\end{figure*}

\begin{figure*}[!htp]
        \centering
            \includegraphics[width=0.41\textwidth]{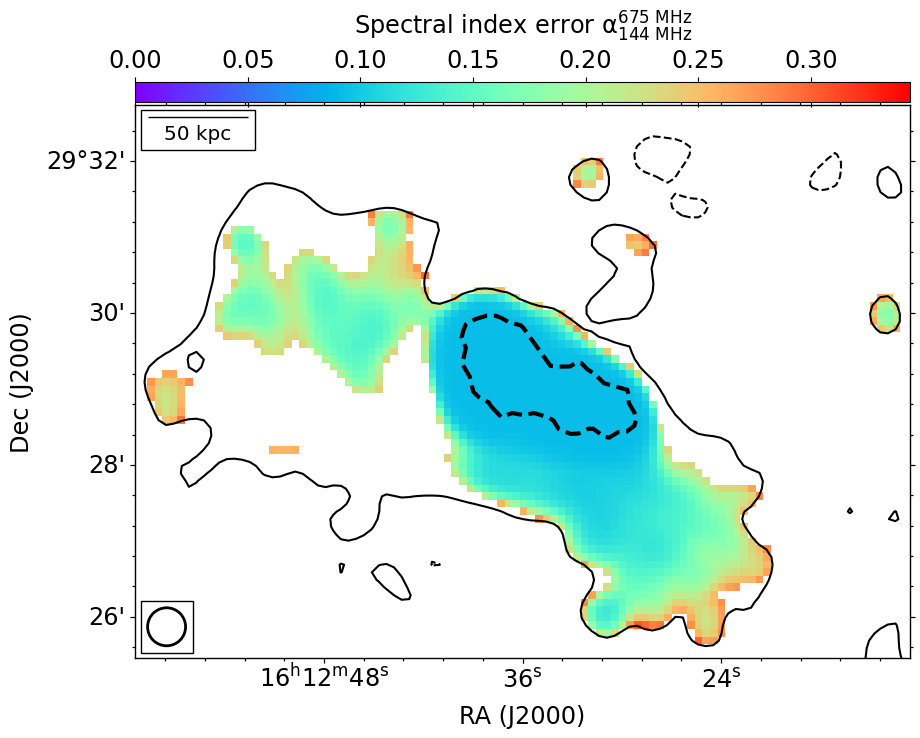}
            \includegraphics[width=0.42\textwidth]{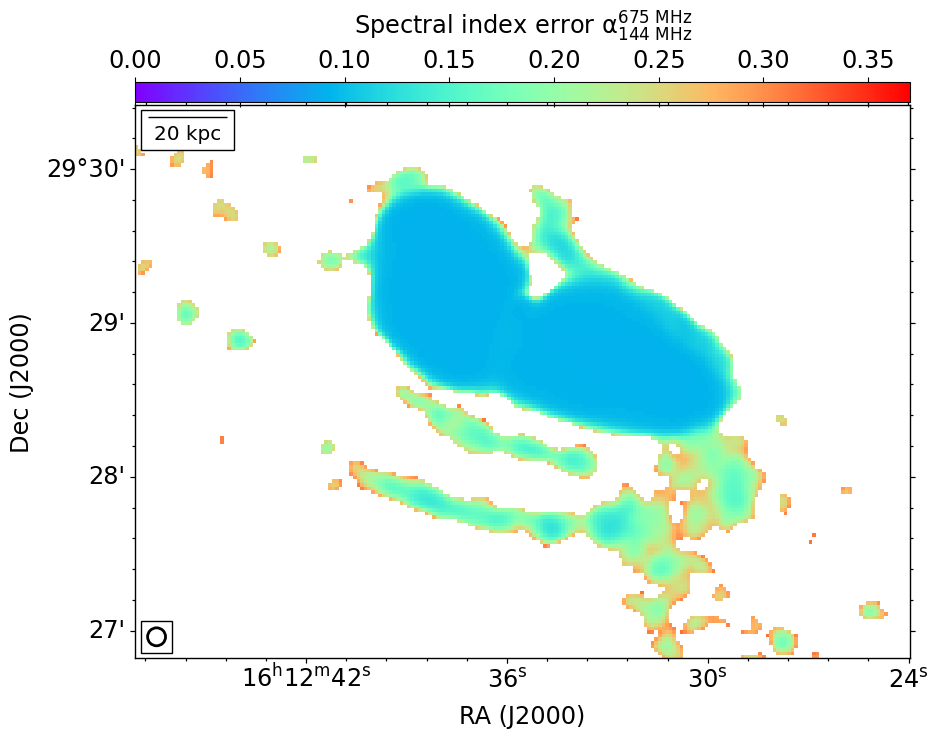}
        \caption{Left: NGC 6086 error value associated with the spectral index map at 30 arcsec between 144-675 MHz. Contours show the LOFAR emission and are drawn at 3$\sigma$. Right: NGC 6086 error value associated with the spectral index map at 7 arcsec between 144-675 MHz. The region is above 3$\sigma$ for all the frequencies. The beam is shown in the bottom-left corner and a reference physical scale is shown in the top-left corner.}
        \label{fig:6086_30_error_map+filaments_error_map}
\end{figure*}

\begin{figure*}[!htp]
        \centering
            \includegraphics[width=0.40\textwidth]{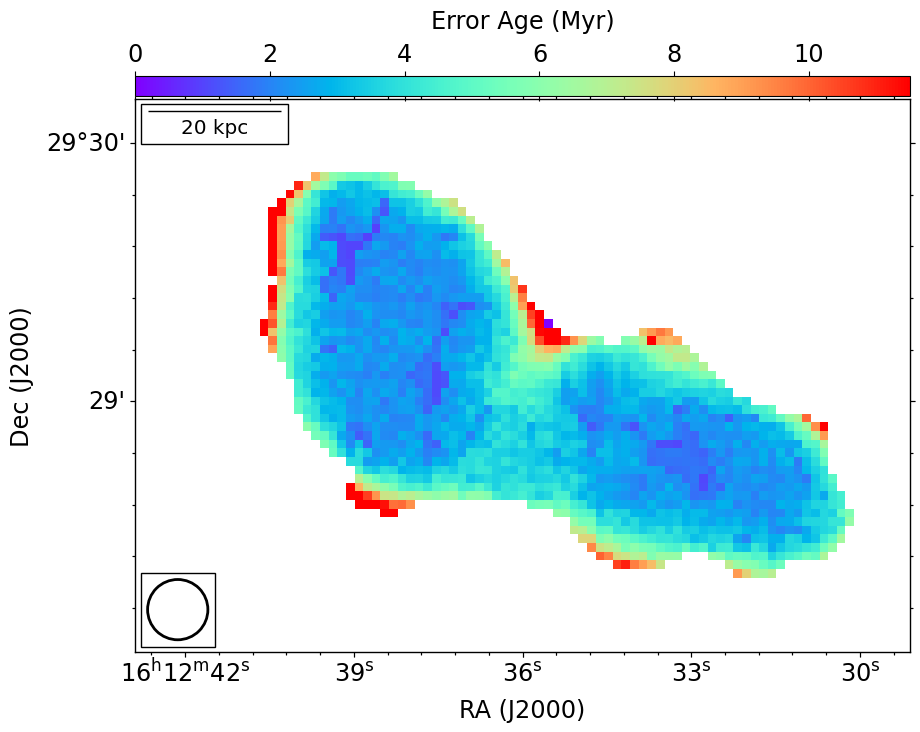}
        \caption{NGC 6086 error value associated with the spectral age map at 14 arcsec between 144-4700 MHz. The region is above 3$\sigma$ for all the frequencies. The beam is shown in the bottom-left corner and a reference physical scale is shown in the top-left corner.}
        \label{fig:age_error_map}
\end{figure*}

\end{document}